\documentclass{aa} 
\usepackage{graphicx}
\usepackage{subcaption}
\usepackage{txfonts}
\usepackage{url}
\usepackage[colorlinks=true,citecolor=blue,linkcolor=magenta,urlcolor=blue]{hyperref}
\usepackage{xcolor}
\usepackage{lscape}

\newcommand{\kms}{km\,s$^{-1}$}

\begin{document} 

\title{Probing the inner Galactic Halo with blue horizontal branch stars:}
\subtitle{Gaia DR3 based catalogue with atmospheric and stellar parameters
\thanks{The catalogue is only available in electronic form at the CDS via anonymous ftp to cdsarc.u-strasbg.fr (130.79.128.5) or via http://cdsweb.u-strasbg.fr/cgi-bin/qcat?J/A+A/}}
\titlerunning{Probing the inner Galactic Halo with blue horizontal branch stars}

\author{R.~Culpan \inst{1}
    \and M.~Dorsch \inst{1,3}
    \and S.~Geier \inst{1}
    \and I.~Pelisoli \inst{2}
    \and U.~Heber \inst{3}
    \and B.~Kub\'atov\'a \inst{4}
    \and M.~Cabezas \inst{4}}

\offprints{R.\,Culpan,\\ \email{rick@culpan.de}}

\institute{Institut f\"ur Physik und Astronomie, Universit\"at Potsdam, Haus 28, Karl-Liebknecht-Str. 24/25, 14476 Potsdam-Golm, Germany
\and Department of Physics, University of Warwick, Coventry, CV4 7AL, UK
\and Dr. Remeis-Sternwarte \& ECAP, Astronomical Institute, FAU Erlangen-Nürnberg, Sternwartstr. 7, 96049 Bamberg, Germany
\and Astronomical Institute AS CR, Fri\v{c}ova 298, 251 65 Ond\v{r}ejov, Czech Republic}

\date{Received 19/10/2023 \ Accepted xx/xx/2024}

% \abstract{}{}{}{}{} 
% 5 {} token are mandatory
 
  \abstract
   {Stars that are found on the blue horizontal-branch (BHB) have evolved from low-mass stars that have completed their core hydrogen burning main sequence stage and have undergone the helium flash at the end of their red-giant phase. They are, hence, very old objects that can be used as markers in studying galactic structure and formation history. The fact that their luminosity is virtually constant at all effective temperatures also makes them good standard candles.}
   {We provide a catalogue of BHB stars with stellar parameters that have been calculated from spectral energy distributions (SED), as constructed from multiple large-scale photometric surveys. In addition, we update our previous, {\em Gaia} Early Data Release 3 catalogue of BHB stars with parallax errors less than 20\% by using the SED results to define the selection criteria. The purpose of these catalogues is to create a set of BHB star candidates with reliable stellar parameters as well as providing a more complete full-sky catalogue with candidate objects found along the whole BHB: from where RR-Lyrae are found on the instability strip to the extreme horizontal branch.}
   {We selected a large dataset of {\em Gaia} Data Release 3 (DR3) objects based only on their position in the colour magnitude diagram, tangential velocity  and parallax errors. Spectral energy distributions were then used to evaluate contamination levels in the dataset and derive optimised data quality acceptance constraints. This allowed us to extend the {\em Gaia} DR3 colour and absolute magnitude criteria further towards the extreme horizontal-branch. The level of contamination found using SED analysis was confirmed by acquiring spectra using the Ond\v{r}ejov Echelle spectrograph attached to the Perek 2m telescope at the Astronomical Institute of the Czech Academy of Sciences.}
   % (resolving power of $R$ = 51600 at 5000$\mathrm{\AA}$) 
   {We present a catalogue of 9,172 Galactic Halo BHB candidate stars with atmospheric and stellar parameters calculated from synthetic SEDs. We also present an extended {\em Gaia} DR3 based catalogue of 22,335 BHB candidate stars with a wider range of effective temperatures and {\em Gaia} DR3 parallax errors of less than 20\%. This represents an increase of 33\% compared to the our 2021 catalogue, with a contamination level of 10\%.}
   {}


\maketitle
\section{Introduction \label{sec:intro}}

The term {\em horizontal branch} (HB) was coined to describe a horizontal structure observed in the colour magnitude diagram (CMD) in globular clusters. The horizontal branch extends from the red clump at the red end through the instability strip where RR-Lyrae are found to the extreme horizontal-branch (EHB) at the blue end \citep{catelan09}. HB stars are what remains of low-metallicity ${\sim}0.8M_{\odot}$ to ${\sim}2.3\,M_{\odot}$ main-sequence stars that have evolved past the helium flash at the end of the red-giant phase. The extent of the HB varies from cluster to cluster which led to the distinction of red HB stars redwards of the instability strip and blue HB stars bluewards. As such these are early type stars that are already billions of years old when they move to the BHB. A more detailed description of BHB stars, their properties and their evolution from main-sequence stars can be found in \citet{moehler01, catelan09} and references therein.

Observationally, the photometry of cooler, A-type, BHB stars places them in the same colour-magnitude space as main-sequence A- and B-type stars. The cooler A-type BHB fall on the bluer side of the instability strip where RR-Lyrae are found \citep{montenegro19}. The hotter B-type BHB stars fall on the redder side of the EHB stars, also called hot subdwarf stars \citep{heber09}. The spectra of BHB stars are similar to A-type subgiant or late B-type main sequence stars. Because BHB stars rotate slowly \citep{behr03, xue08}, their spectral lines are usually sharper than main-sequence A- and B-type stars. 

\begin{figure}
  \centering
  \includegraphics[width=\hsize]{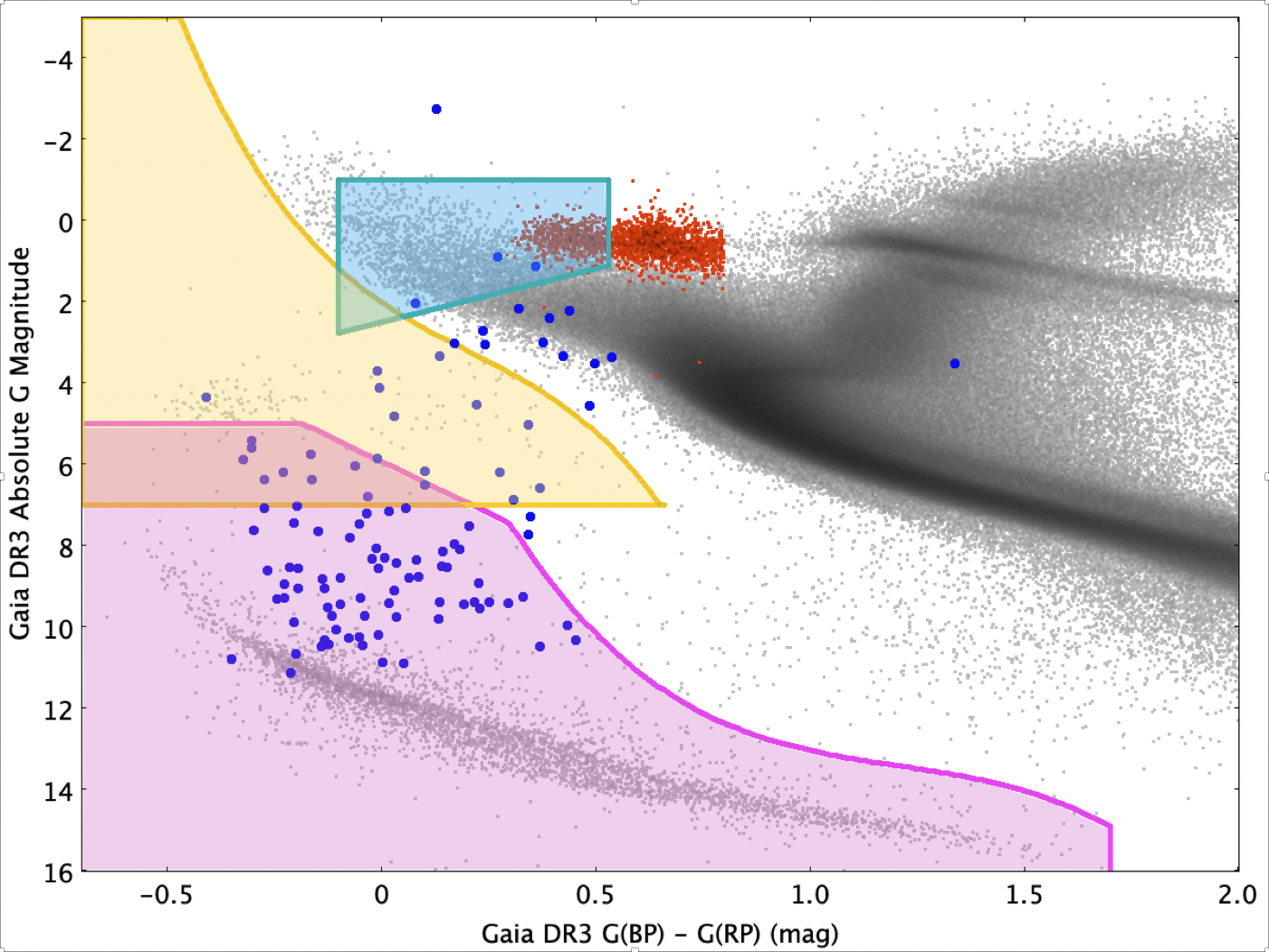}
 \caption{The Gaia DR3 colour magnitude diagram for 30,000 randomly selected objects with parallax errors less than 10\% and tangential velocities greater than 145 \kms\ (grey dots); BHB CMD selection region from \citet{culpan21} (turquoise shading); a selection of RR-Lyrae from Gaia DR3 \citet{clementini22} (red dots); hot-subdwarfs CMD selection region from \citet{culpan22} (yellow shading); white dwarfs CMD selection region from \citet{gentile21} (pink shading); (pre-)ELM white dwarfs from \citet{pelisoli19} (blue circles).}
  \label{star_types}
  \end{figure}

These classifications rely on an accurate and precise determination of colour and absolute magnitude. {\em Gaia} has enabled immense progress to be made in this field but even with {\em Gaia} DR3 the uncertainties in parallax, colour, apparent magnitude, and proper motions still lead to a degree of misplacement of some stars in the CMD causing contamination in catalogues that are based on {\em Gaia} DR3 data alone. This presents the additional challenge of removing such contaminants. The main categories of stars that plot closest to the BHB in the colour-magnitude diagram, as illustrated in Figure~\ref{star_types}, are hot subdwarfs \citep[see e.g.][]{culpan22}, extremely low mass (ELM) white-dwarfs and pre-ELM \citep[as compiled by][]{pelisoli19}, and RR-Lyrae \citep[see ][]{eyer23}. Further contamination is caused by blue straggler stars, normal main-sequence stars that have been rejuvenated through the accretion of additional hydrogen \citep{mccrea64,shara97,chen09,ekanayake18}. They follow the part of the main-sequence that is brighter and bluer than the main-sequence turnoff of the Galactic Halo, i.\ e.\ where normal main-sequence A-type and B-type would no longer be present in older stellar populations. The brightest blue stragglers can be up to three  magnitudes brighter than the main-sequence turnoff \citep{rain21}. 

Blue stragglers, because of their somewhat higher gravities, can be distinguished from A-type BHB stars spectroscopically \citep[][and references therein]{xue08}, while this is difficult for B-type stars because of their similar gravities \citep{hunger87,raddi21}. Main-sequence stars of spectral type A- and B- \citep[see e.g.][]{silva11,raddi21,liu23} are also found in this region of the colour-magnitude diagram but are not expected in the Galactic Halo because their expected main-sequence lifetime is shorter than the age of this region. However, low numbers of younger runaway main-sequence stars that have been ejected from the Galactic Disc are found in this region.

Here we present a new release of the {\em Gaia} DR3 based catalogue of halo BHB stars that is more complete than the original {\em Gaia} Early Data Release 3 (EDR3) based version \citep{culpan21} in crowded regions, particularly close to the Galactic Plane, and at the bluer end of the HB where late B-type BHB stars are expected. The catalogue of Galactic Halo BHB candidate stars with stellar parameters derived from their spectral energy distributions (SEDs) is a subset of the newly released {\em Gaia} DR3 based catalogue. The SEDs have been generated using photometric data from multiple large-scale surveys.  We have used the SED results to modify the {\em Gaia} DR3 selection criteria to populate the catalogue. 
They were further combined with newly acquired spectra to estimate the level of contamination in the catalogue. As such, the process to define the BHB selection criteria has been an iterative one. Only the final result shall be described in this paper.

The contents of this paper are as follows: In Section \ref{sect:SED} we explain the generation of stellar parameters using SEDs and apply this method to the 16,794 BHB candidates in the parallax selection of \citet{culpan21} and use the results to revise and extend the {\em Gaia} DR3 CMD selection criteria. Section \ref{sect:revision} describes how we selected and verified the applicability of various astrometric, photometric, and blending quality criteria when applied to our particular {\em Gaia} DR3 data set. In Section \ref{sect:contamination} we estimate the contamination levels and completeness of the {\em Gaia} DR3 catalogue using the stellar parameters from the SEDs as well as acquired spectra. Section \ref{sect:sky} covers the apparent magnitude range, distances and sky coverage of the {\em Gaia} DR3 catalogue. In Section \ref{sect:summary} we summarise our findings and draw our conclusions.

%--------------------------------------------------------------------
\section{Spectral energy distribution analysis of BHB candidates}
\label{sect:SED}

We used all of the 16,794 BHB star candidates with parallax errors < 20\% found by \citet{culpan21} as the starting point for the SED analysis. The SED of a star is defined as the observed spectral flux density as a function of wavelength. This involved retrieving the photometric data from 66 surveys, where available (see Appendix \ref{appendix:photometry}). Photometric data have only been used in SED generation if they had no error flags set and thus conformed to all the quality criteria for the survey from which they came.

\begin{figure}
  \centering
  \includegraphics[width=\hsize]{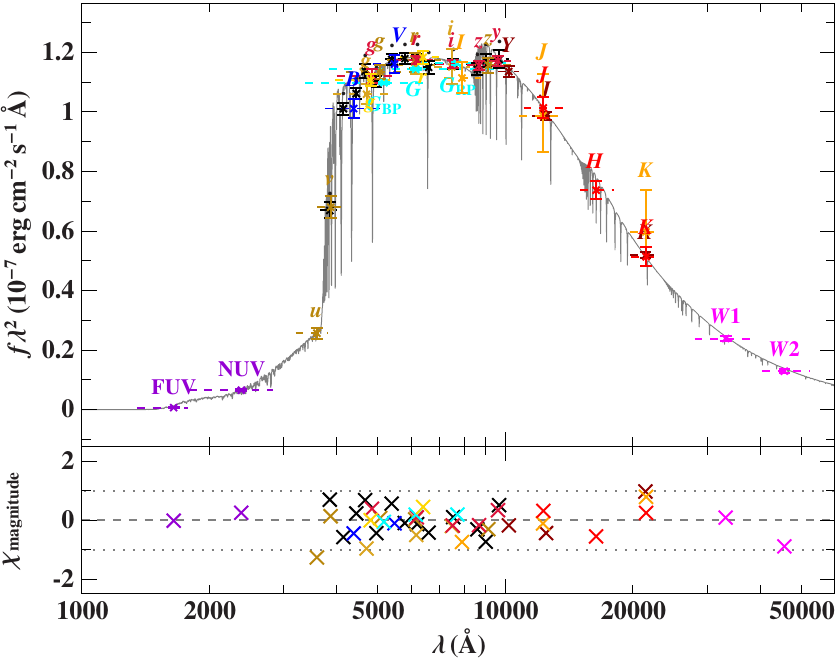} % SED_Gaia_DR3_6271805644154197504.png
 \caption{An example of the results of a spectral energy distribution generated for the BHB candidate star {\em Gaia} DR3 6305433829332391168 with the photometric data labelled. The grey line is the best-fit SED solution for the available photometric data. The bottom panel showing the residuals between the best fit solution and the photometric data.}
  \label{sed_example}
  \end{figure}

Grids of synthetic SEDs were calculated with an amended version of the ATLAS12 \citep{kurucz96, irrgang18} code. 
A detailed description of the SED fit method used here can be found in \citet{heber18} and \citet{latour23}. An example is shown in Fig.~\ref{sed_example} and Table \ref{table:B1} lists photometic data used for the full sample. 
% \citet{irrgang18}
Free parameters in this $\chi^2$ fit were the star's effective temperature ($T_\mathrm{eff}$), surface gravity ($\log g$), angular diameter ($\Theta$), as well as the colour excess caused by interstellar extinction. The latter was treated using the extinction law of \citet{Fitzpatrick2019}  for a reddening parameter of $R_{55}=3.02$, the mean value for the Milky Way. Because the metallicity could not be determined from the SED alone, we assumed $\mathrm{[Fe/H]} = -1.7$, a standard value for cool  BHB stars in the Galactic field \citep{behr03}. Stars hotter than the Grundahl jump  \cite[11500\,K,][]{Grundahl1999} were assigned solar metallicity because their metal abundances are increased as a result of diffusion. 

\citet{latour23} successfully used this approach for BHB stars in globular clusters, as shown in their fig.\ 7. 
The angular diameter allows a measurement of the stellar radius $R$ when combined with a distance measurement: $R = \Theta/(2\varpi)$, where $\varpi$ is the parallax. Here, we used the {\em Gaia} DR3 parallax measurements, corrected for zero-point offsets following \citet{lindegren20}.

\begin{figure}
  \centering
  \includegraphics[width=\hsize]{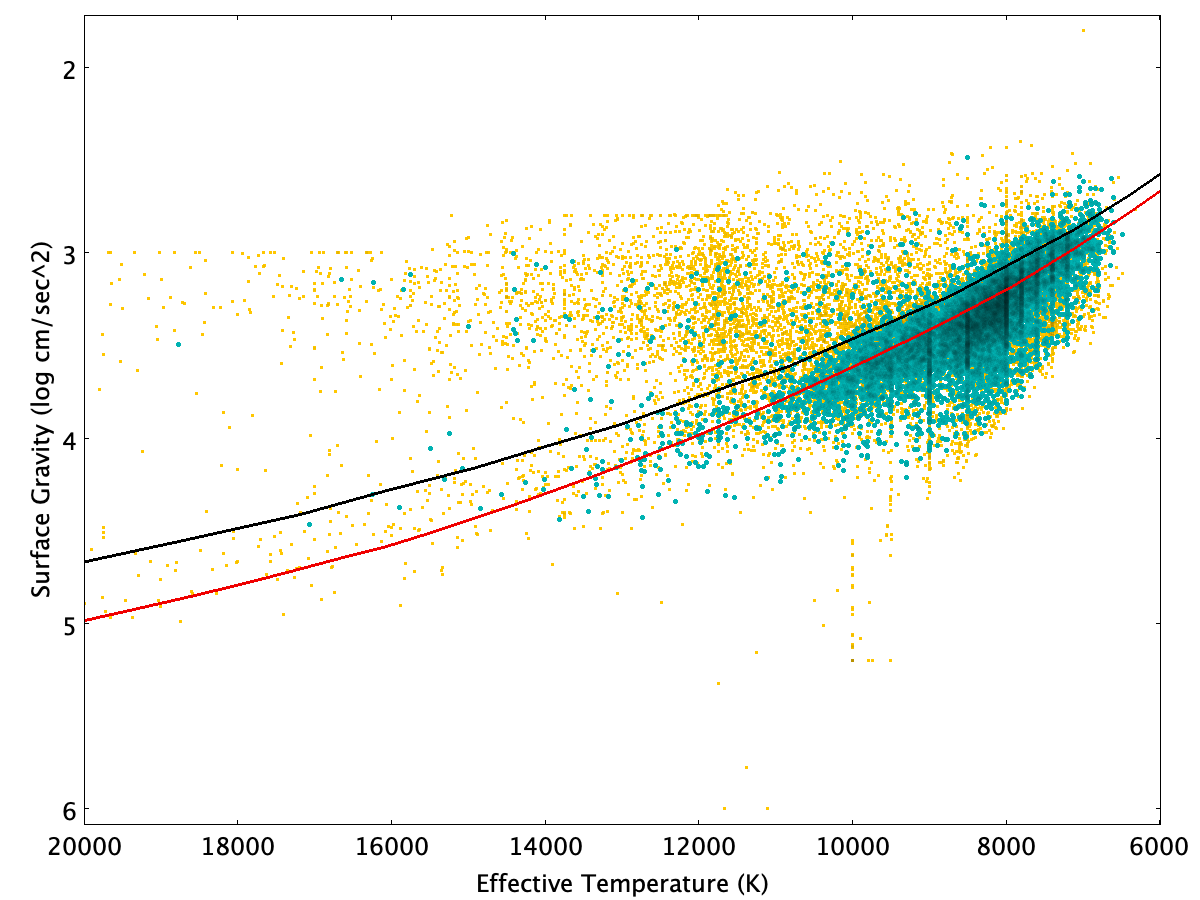}
  \includegraphics[width=\hsize]{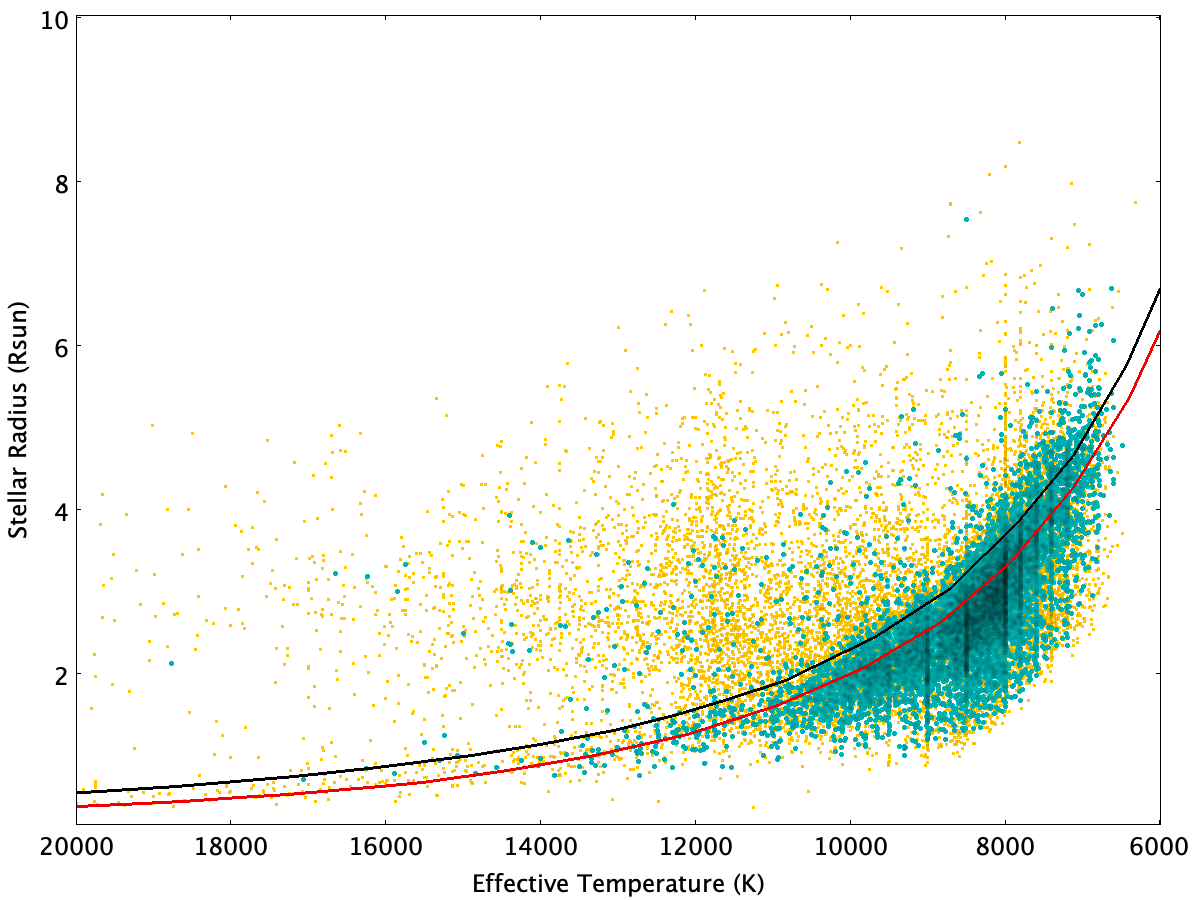}
 \caption{The stellar parameters calculated by the synthetic SEDs from photometric data. All SED results are shown as yellow circles, the SED results for objects with full wavelength range photometric data available are shown as turquoise triangles. The Red line is the BaSTI ZAHB line. The black line is the BaSTI TAHB.  Upper panel:$T_\mathrm{eff}$ versus $\log g$; Lower panel: $T_\mathrm{eff}$ versus $R$.}
  \label{SED_QC_results}
  \end{figure}

\begin{figure}
  \centering
  \includegraphics[width=\hsize]{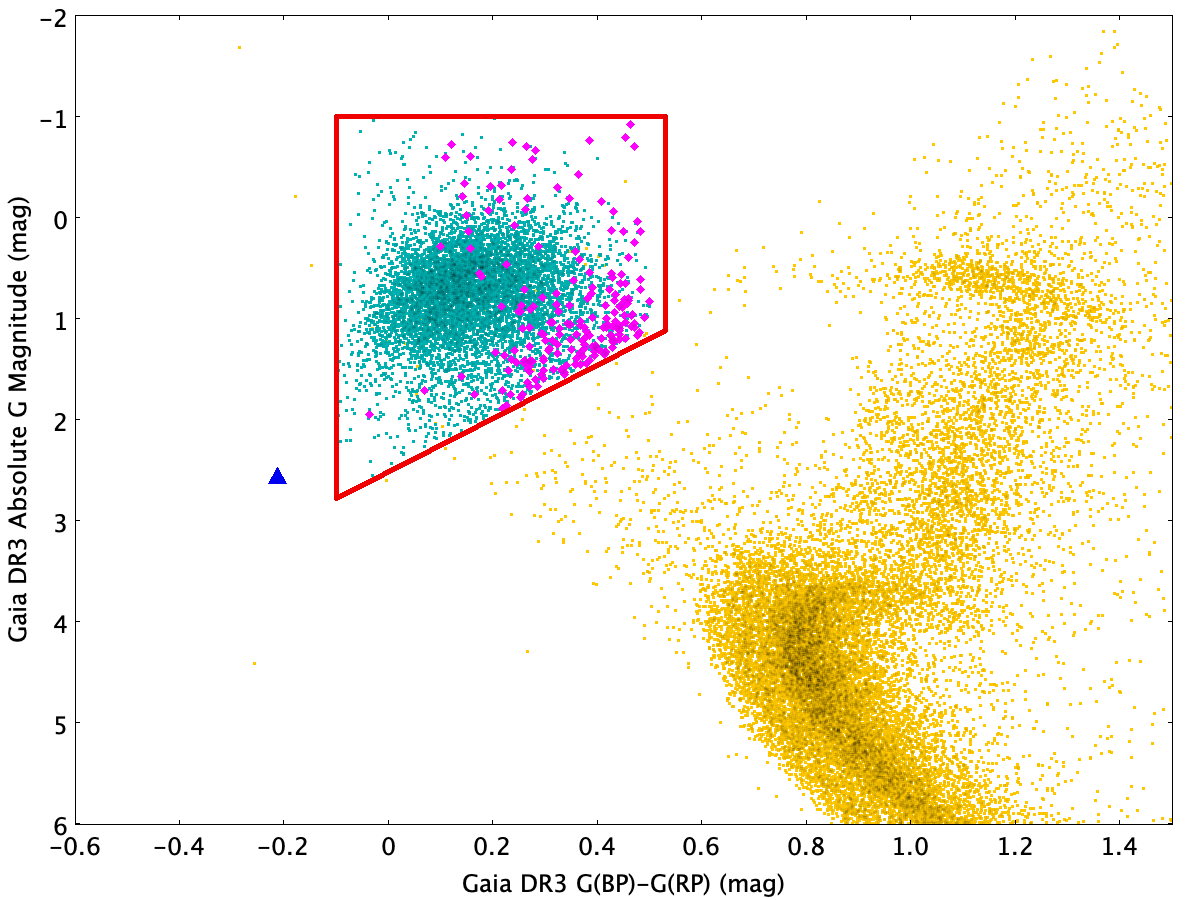}
 \caption{The {\em Gaia} DR3 CMD showing a sample of stars with tangential velocities greater than 145 \kms (yellow dots), \citet{culpan21} objects with full range SEDs (cyan dots), \citet{culpan21} objects with full range SEDs and modelled reddening greater than 0.2 mag (magenta squares) and the known BHB Feige 86 possessing $T_\mathrm{eff}$ around 15,000K \citep{nemeth17} (blue triangle) which lies outside the {\em Gaia} DR3 CMD BHB region from \citet{culpan21}.}
  \label{culpan21_feige86}
  \end{figure}

We found that the SEDs for stars where photometric data were available over the full range of wavelengths (from ultra-violet to infra-red) gave the most reliable results on the $T_\mathrm{eff}$ versus $\log g$ and the $T_\mathrm{eff}$ versus $R$ plots with the BHB candidates plotting close to the BaSTI zero-age horizontal-branch (ZAHB) and the terminal-age horizontal-branch (TAHB) lines as defined in \citet{pietrinferni04} and updated in  \citet{hidalgo18}(see our figure~\ref{SED_QC_results}). The BHB candidate stars with full range wavelength photometry available plotted in a well-defined cloud in the BHB region of the {\em Gaia} DR3 colour-magnitude diagram, (see Figure~\ref{culpan21_feige86}). We defined the full range of wavelengths as having data available from less than 2000\,$\mathrm{\AA}$ to over 8000\,$\mathrm{\AA}$.

\begin{figure}
  \centering
  \includegraphics[width=\hsize]{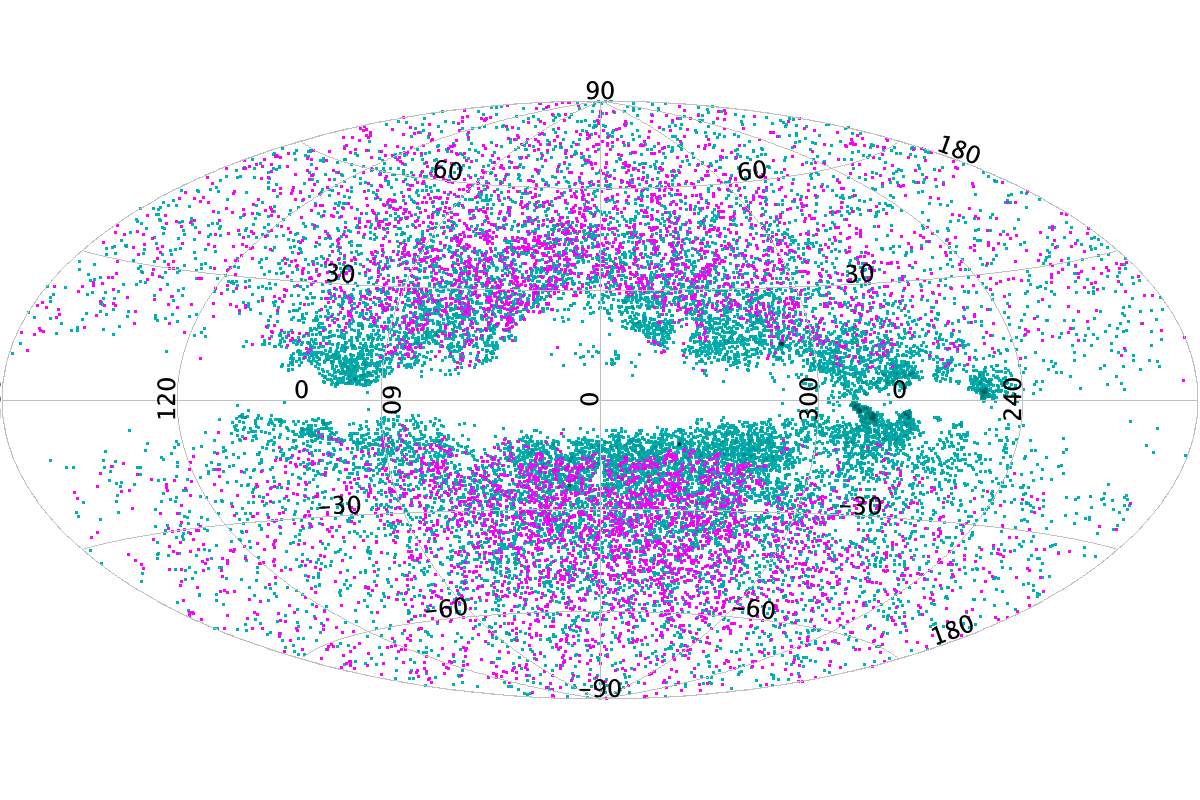}
 \caption{Sky distribution of {\em Gaia} DR3 BHB candidate stars (cyan dots) and the catalogue of BHB stars with stellar parameters based on SEDs using full photometric range data (magenta dots).}
  \label{all_sky}
  \end{figure}

In order to ensure that no selection effects were distorting our results we verified that the BHB candidates with full range photometric data were not clustered within a particular colour, parallax, or apparent magnitude range. Full range photometric data was available for Galactic halo objects across all of the sky including regions of high apparent stellar density (see Figure~\ref{all_sky}) giving us confidence that our results from the SED analysis were applicable to the entire BHB candidate catalogue.

We plotted the SED results for  $T_\mathrm{eff}$ versus $\log g$ and $T_\mathrm{eff}$ versus stellar radius to identify which of the BHB candidate stars with photometric data over a wide wavelength range plotted between the ZAHB and the TAHB lines (see Figure~\ref{SED_QC_results}). We found that the BHB candidate CMD selection of \cite{culpan21} was largely restricting the upper limit of the $T_\mathrm{eff}$, thus excluding hotter BHB stars with B-type spectra. In particular we noted that the known BHB star Feige 86 with $T_\mathrm{eff}$ around 15,000K \citep{nemeth17} was not present in the \citet{culpan21} catalogue (see Figure~\ref{culpan21_feige86}).

We found that the interstellar reddening that was calculated for the SED generation was a critical issue. When we included the objects that had reddening values greater than 0.2 mag in the candidate selection we found a greatly increased scatter in the CMD plots (see Figure~\ref{culpan21_feige86})

We also found that the main-sequence contamination levels increased where the BHB candidates with high reddening overlapped with the main-sequence in the CMD. In this way we used the reddening as calculated in the SED generation to define the {\em Gaia} DR3 CMD selection criteria and the upper limit on absolute magnitude was taken to be 0.2 mag extinction line when plotted on the {\em Gaia} DR3 CMD.

\section{Revising the BHB star candidate catalogue from {\em Gaia} DR3 using the SED results}
\label{sect:revision}

\begin{figure}
  \centering
  \includegraphics[width=\hsize]{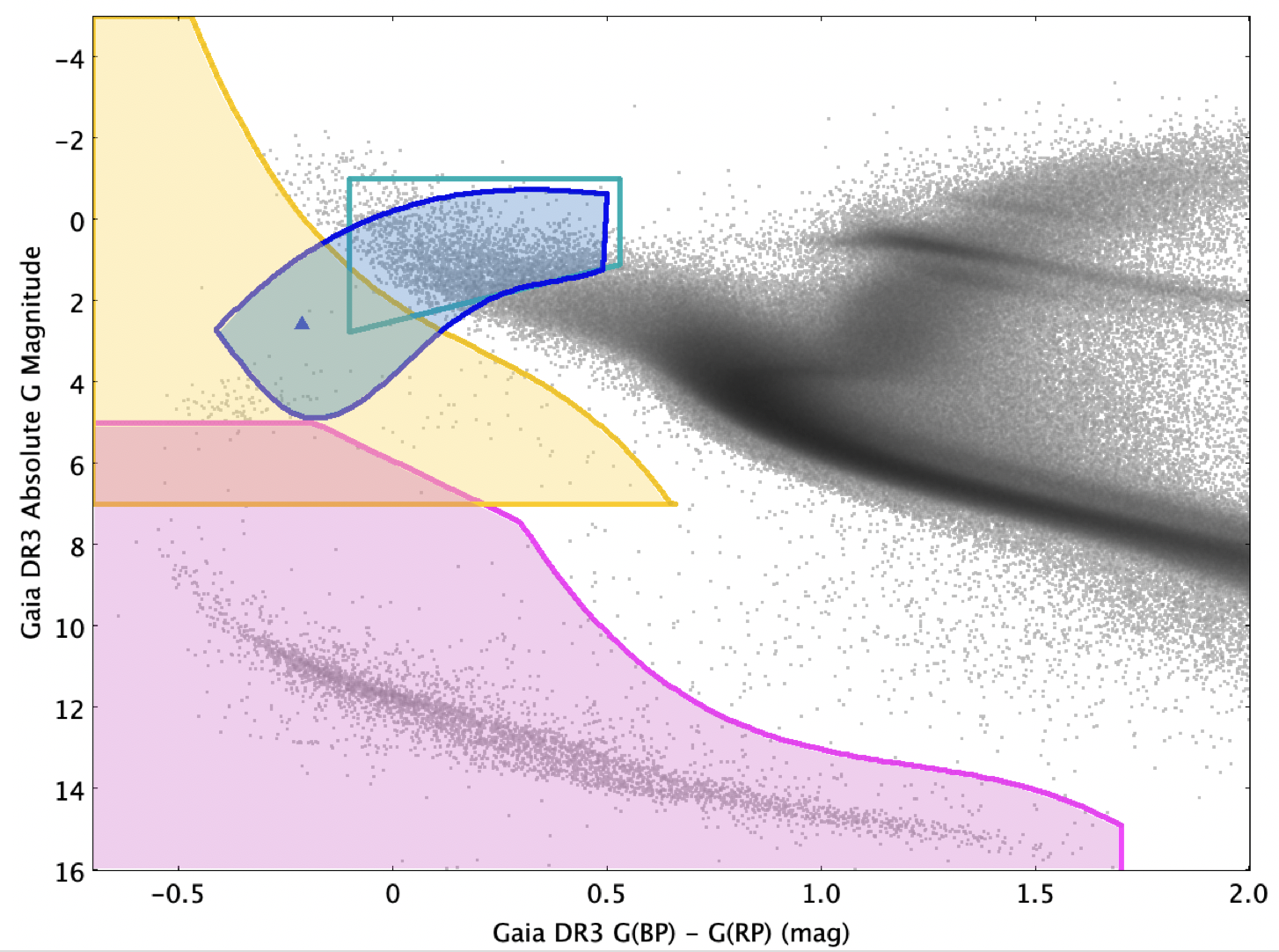}
 \caption{The Gaia DR3 colour magnitude diagram for 30,000 randomly selected objects with parallax errors less than 10\% and tangential velocities greater than 145 \kms\ (grey dots); BHB CMD selection region from \citet{culpan21} (turquoise outline); extended BHB CMD selection region (blue shading); hot-subdwarfs CMD selection region from \citet{culpan22} (yellow shading); white dwarfs CMD selection region from \citet{gentile21} (pink shading); Feige 86 (blue triangle).}
  \label{new_star_types}
  \end{figure}

In order to raise the upper limit on the $T_\mathrm{eff}$ we extended the {\em Gaia} DR3 CMD search criteria towards the cluster of hot subdwarfs found in \citet{culpan22}. The resulting overlap (see Figure~\ref{new_star_types}) means that 643 of the BHB candidates found are also hot subdwarf candidates as found in \citet{culpan22}.

Based on {\em Gaia} DR3 data we calculated the absolute $G$ magnitude ($G_\mathrm{abs}$), tangential velocity ($V_T$), proper motion error ($\sigma_\mathrm{pm}$), and tangential velocity error ($\sigma_{V_{T}}$) as:
\begin{equation}
\label{eq:gabs}
    G_\mathrm{abs} = \verb!phot_g_mean_mag! + 5 + 5\,\log_{10}(\verb!parallax!/1000),
\end{equation}\vspace{-10pt}
\begin{equation}
\label{eq:vt}
    V_T = 4.74 \cdot \texttt{pm}/\texttt{parallax},
\end{equation}\vspace{-10pt}
\begin{equation}
    \sigma_\mathrm{pm} = \frac{\sqrt{(\sigma_\mathrm{pmra}\,\texttt{pmra})^2 + (\sigma_\mathrm{pmdec}\,\texttt{pmdec})^2}}{\texttt{pm}},
\end{equation}\vspace{-10pt}
\begin{equation}
    \sigma_{V_{T}} = V_T \sqrt{\frac{\sigma_\mathrm{pm}}{\texttt{pm}} + \frac{\sigma_\mathrm{parallax}}{\texttt{parallax}}}.
\end{equation}

The {\em Gaia} DR3 CMD search criteria were also revised to follow the BaSTI theoretical $G_\mathrm{abs}$ versus $(G_{\rm BP} - G_{\rm RP})$ lines relating to the 0 mag and 0.2 mag reddening \citep{pietrinferni04,hidalgo18}. The actual selection criteria used (see Table 1 and Appendix C) were somewhat wider than the 0 mag and 0.2 mag reddening lines (see Figure~\ref{results}) to allow for errors in the calculated absolute magnitude arising from uncertainties of up to 20\% in the {\em Gaia} DR3 parallax measurements.

\subsection{Determination of the {\em Gaia} DR3 data quality filtering criteria}

Since the release of {\em Gaia} DR3 many studies were made using many different quality selection criteria. We have used the stellar parameters generated using SEDs to quantify the effect that different criteria have on the levels of contamination found in this catalogue and thus support our own quality criteria selection. We have defined contaminants as being stars that do not plot in the BHB region of the $T_\mathrm{eff}$ versus $R$ and the $T_\mathrm{eff}$ versus $\log g$ plots. It should be noted that our quality criteria selection findings apply to our BHB star catalogue and may not necessarily be transferable to other situations using {\em Gaia} DR3 data.

The selection of the parallax quality criterion was examined by looking at the scatter observed in the CMD as the parallax error cutoff was varied. We observed an increase in scatter in the cloud of BHB candidates on the CMD as the parallax error cutoff was increased but there was no obvious point (e.\,g.\ a sudden increase or drop) at which a best value could be justified. We have, thus, continued to use the parallax error < 20\% as the parallax quality criterion for the sake of consistency with the \citet{culpan21} and \citet{culpan22} catalogues. Relaxing the parallax quality criterion will consequently increase the possible errors of the absolute magnitude and the tangential velocity calculations (see Equations \ref{eq:gabs} and \ref{eq:vt}) which will have the effect of increasing contamination levels within the catalogue.

The effect of applying the astrometric quality selection criterion used in \citet{culpan21}, \verb!astrometric_sigma5d_max! < 1.5, which was taken from \citet{lindegren20} was compared to the \verb!fidelity_v2! criterion from \citet{rybizki22}. We found that \citet{rybizki22} removed an additional 92 sources from the parallax error < 20\% candidates compared to \citet{lindegren20}. This represents less than 1\% of the candidates. Examination of the SEDs for these 92 objects showed no benefit in their removal in terms of reduction of contamination levels. We have retained the use of the \citet{lindegren20} criterion.

We applied the photometry quality criterion from \citet[][section 9.4]{riello21} that calculates the \verb!astrometric_excess_noise_corrected! = $C^{*}$ and suggests a fitted scatter line
\begin{equation}
\label{eq:C}
  |C^{*}| < N \cdot \sigma_{C^{*}}
\end{equation}
Varying the value of $N$ in Equation \ref{eq:C} between 1 and 5 merely changes the number of sources selected by 25 (in a population of over 13,000) and was, as such, deemed to be of negligible significance. We have retained the value of 5 as recommended in \citet[][section 9.4]{riello21}. 

We compared the crowded region filter from \citet{culpan22} to the blending criterion from \citet[][section 9.3]{riello21}. The \citet{culpan22} criterion requires that the BHB candidate's $G$ flux must make up at least 70\% of the $G$ flux detected within a 5 arcsec radius around that candidate. The blending criterion from \citet{riello21} is defined as the number of transits where the $G_{BP}$ or $G_{RP}$ photometry is likely  blended with nearby sources, compared to the total number of transits used for the photometry measurements. We found that the (\verb!beta! < 0.1) criterion from \citet{riello21} removed an additional 8,639 BHB candidates compared to the \citet{culpan22} criterion. Once again, the contamination level found from the SED analysis was not reduced by applying the more restrictive criterion. Thus, we retained the crowded region filter from \citet{culpan22}.

The application of the data quality filters as described above to the {\em Gaia} DR3 data set resulted in 23,415 BHB candidates (see Table 1).

\subsection{Sources of catalogue contamination}

There is a compromise to be met between completeness and contamination in our {\em Gaia} DR3 based catalogue of BHB stars. The contamination levels found may be reduced by placing tighter quality and selection criteria but this will also reduce the number of candidate stars found. We considered two types of contaminants to our catalogue.

The first type of contaminant is given by stars that should not plot in the same {\em Gaia} DR3 colour magnitude space as BHB stars but do so because of erroneous {\em Gaia} DR3 measurements. The main contaminants in this category will be those objects that plot closest to the BHB in {\em Gaia} DR3 colour magnitude space. The main contaminants of this type are hot subdwarfs, pre-ELM white dwarfs, and RR-Lyrae stars. The closer the true proximity of these objects to the BHB region and the larger the population the greater the expected contamination. 

\begin{figure}
  \centering
  \includegraphics[width=\hsize]{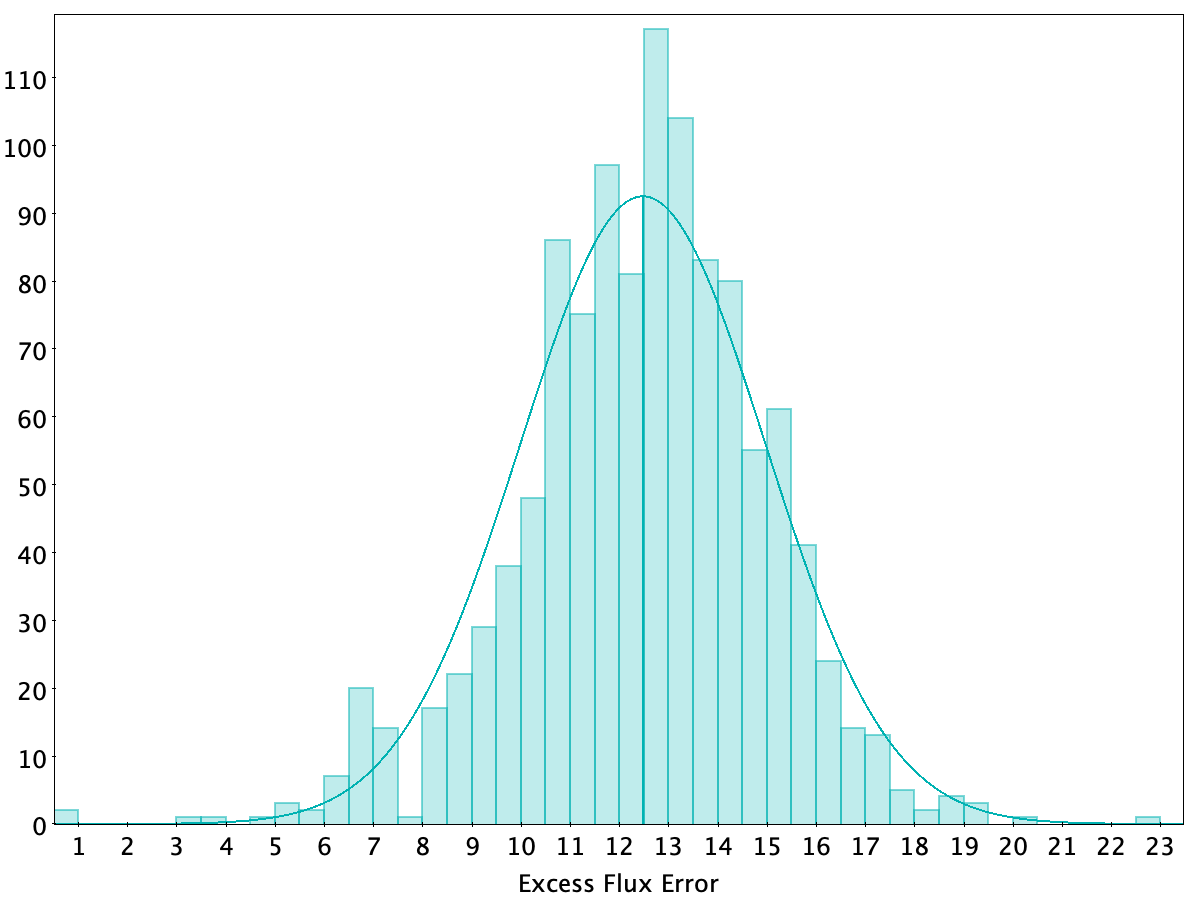}
 \caption{The distribution of excess flux error for known SIMBAD and {\em Gaia} DR3 RR Lyrae that are found within our catalogue of BHB candidates. A best-fit Gaussian distribution (cyan line) as been superimposed.}
  \label{rrlyrae_efe}
  \end{figure}

The most common contaminants of this type are RR-Lyrae stars. The low temperature end of the BHB meets the instability strip where RR-Lyrae are found on the CMD (see Figure \ref{star_types}). 
Our aim is to generate a catalogue of non-intrinsically variable stars that lie between the instability strip and the EHB, making these variable stars a significant contaminant in the catalogue. We cross-matched our BHB candidate stars with the SIMBAD database and the {\em Gaia} DR3 variability RR-Lyrae catalogue \citep{eyer23, clementini22} to find known RR-Lyrae. Of our BHB candidates we found 1,004 known RR-Lyrae from the SIMBAD database and 1,136 in the Gaia catalogue. It should be noted that only a small fraction of the \citet{clementini22} RR-Lyrae that satisfy the parallax quality criterion, the tangential velocity selection criterion and the BHB CMD selection criteria. We then calculated the excess flux error as defined in \citet{gentile21} and \citet{culpan22} for all BHB candidates. The excess flux error is calculated by comparing the \verb!phot_g_flux_error! of each candidate object to the median $G$ flux error of 500 similar objects in terms of colour (\verb!bp_rp!), $G$ flux ($\log_{10}$~\verb!phot_g_mean_flux!), and the number of observations (\verb!phot_g_n_obs!) taken from the full {\em Gaia} DR3 catalogue. The flux error method is sensitive to any variable stars within the limits of the {\em Gaia} exposure time and the timescale of repeat observations. We found that over 90\% of the known RR-Lyrae had an excess flux error greater than 6.0 (see Figure~\ref{rrlyrae_efe}). This was then used as a criterion for the removal of a further 189 candidate variable stars that are not found among those that are known in SIMBAD or in the {\em Gaia} DR3 variability RR-Lyrae catalogue.

The fact that the remaining BHB candidates that were not identified as being variable have an average excess flux error of 0.4 demonstrates that excess flux error is an effective differentiator for variability.

The second type of contaminant is stars that are not BHB stars but correctly plot in the same region in {\em Gaia} DR3 colour magnitude space. This category includes young A-type and B-type main-sequence stars and blue stragglers. We have, as previously stated, focussed on the Galactic Halo to remove contamination from main-sequence A- and B-type stars by setting a lower limit on the tangential velocity. This criterion will, however, not filter out high proper motion disk objects. Furthermore, a 20\% error on parallax will lead to a scatter in tangential velocities of up to 40 km\,s$^{-1}$ around a mean value allowing further contamination from main-sequence disk objects with higher errors on their parallax measurement. It should be noted that the tangential velocity that is calculated from $\texttt{pmra}$ and $\texttt{pmdec}$ is calculated in the heliocentric reference frame and not the Galaxy centered reference frame. Stars in the Galactic Halo may have a high radial velocity along the line of sight and hence a low tangential velocity in the heliocentric reference frame. Such stars may, thus, not be selected. This effect is more prevalent in the direction of motion of the heliocentric system.

We applied the extended {\em Gaia} DR3 CMD selection criteria for BHB stars while applying the astrometric, photometric, crowded region, and RR-Lyrae rejection criteria given above (see Table~\ref{table:2}). This resulted in 22,335 BHB candidate stars.

\begin{table*}[h!]
\caption{Table of criteria to select the BHB candidate stars.}
\label{table:2}
\centering
\begin{tabular}{c} 
 \hline
 \hline
 \noalign{\smallskip}
 \it 1. {\em Gaia} DR3 CMD selection (see Appendix C): 30,228 objects\\
 \noalign{\smallskip}
 \hline
 \noalign{\smallskip}
 $\verb!parallax! > 0$ \\ 
 $\verb!parallax_over_error! > 5$ \\ [0.5ex]
 $-0.4 < (G_{\rm BP} - G_{\rm RP}) < 0.5$ \\
 $G_{\rm abs} < 138.07 (G_{\rm BP} - G_{\rm RP})^6 - 153.85 (G_{\rm BP} - G_{\rm RP})^5 - 40.727 (G_{\rm BP} - G_{\rm RP})^4 + 73.368 (G_{\rm BP} - G_{\rm RP})^3$ \\
 $-7.4054 (G_{\rm BP} - G_{\rm RP})^2) - 9.5575 G_{\rm BP} - G_{\rm RP} + 3.8459$ \\
 $G_{\rm abs} > -3.2382 (G_{\rm BP} - G_{\rm RP})^3 + 7.1259 (G_{\rm BP} - G_{\rm RP})^2 - 3.583 (G_{\rm BP} - G_{\rm RP}) - 0.2$ \\ [0.5ex]
 $v_t >= 145 km/s$ \\ [0.5ex]
 \noalign{\smallskip}
 \hline
 \hline
 \noalign{\smallskip}
 \it 2. Photometric quality selection criteria: 30,088 objects\\
 \noalign{\smallskip}
 \hline
 \noalign{\smallskip}
 $|C^{*}| < 5. \sigma_{c^{*}}$ \\ [0.5ex]
 \noalign{\smallskip}
 \hline
 \hline
 \noalign{\smallskip}
 \it 3. Blended object rejection criterion: 23,619 objects\\
 \noalign{\smallskip}
 \hline
 \noalign{\smallskip}
 BHB candidate flux fraction from 5 arcsec radius > 0.7 \\ [0.5ex]
 \noalign{\smallskip}
 \hline
 \hline
 \noalign{\smallskip}
 \it 4. RR-Lyrae removal criterion: 22,335 objects \\
 \noalign{\smallskip}
 \hline
 \noalign{\smallskip}
 Objects listed as RR-Lyrae in Simbad \\
 Objects considered RR-Lyra candidates in Gaia DR3 \\ [0.5ex]
 $\verb!excess_flux_error! > 6.0$ \\ [0.5ex]
 \noalign{\smallskip}
 \hline
 \hline
 \noalign{\smallskip}
 \it 5. BHB candidates with wide wavelength range photometric data: 10,604 objects \\
 \noalign{\smallskip}
 \hline
 \noalign{\smallskip}
 from 2000 \AA\ to 8000 \AA \\ [0.5ex]
 \noalign{\smallskip}
 \hline
 \hline
 \noalign{\smallskip}
 \it 6. SEDs calculated without error flags: 10,222 objects \\
 \noalign{\smallskip}
 \hline
 \noalign{\smallskip}
 At least 5 photometry measurements used for the fit \\
 Maximum 20\% uncertainty in radius \\
 Maximum 0.3dex uncertainty in log(L)\\ [0.5ex]
 \noalign{\smallskip}
 \hline
 \hline
 \noalign{\smallskip}
 \it 5. SED generated effective temperature and surface gravity plot within BHB region of Kiel diagram: 9,172 objects \\
 \noalign{\smallskip}
 \hline
 \noalign{\smallskip}
 $ \log_{10} (g) < 4.4157  \log_{10}(T_\mathrm{eff}) - 13.720$ \\
 $ \log_{10} (g) > 4.0150  \log_{10}(T_\mathrm{eff}) - 12.742$ \\ [0.5ex]
 \noalign{\smallskip}
 \hline
 \hline
\end{tabular}
%\noalign{\smallskip}
\end{table*}

\section{Catalogue contamination estimation}
\label{sect:contamination}

In order to estimate the levels of contamination in the {\em Gaia} DR3 catalogue of BHB stars we used two independent methods that both were not dependent on {\em Gaia} DR3 data: the SED method as described in Section \ref{sect:revision} as well as the analysis of spectra acquired specifically for this project using the Astronomical Institute of the Czech Academy of Sciences’ Perek 2m telescope located at Ond\v{r}ejov in the Czech Republic.

\subsection{Contamination estimation using SED results}
\label{sect:cont_sed}

As previously stated in Section \ref{sect:SED}, the BHB candidate stars with a full wavelength range of photometric data available give the most reliable SEDs. Full range photometric data were used in generating the SED for 10,604 of the 22,335 (53\%) BHB candidates found (see Table 1). As we found no apparent selection effects with regard to {\em Gaia} DR3 colour, parallax, or apparent magnitude we consider the BHB candidates with full range photometric data to be representative of the catalogue as a whole.

\begin{figure}
  \centering
  \includegraphics[width=\hsize]{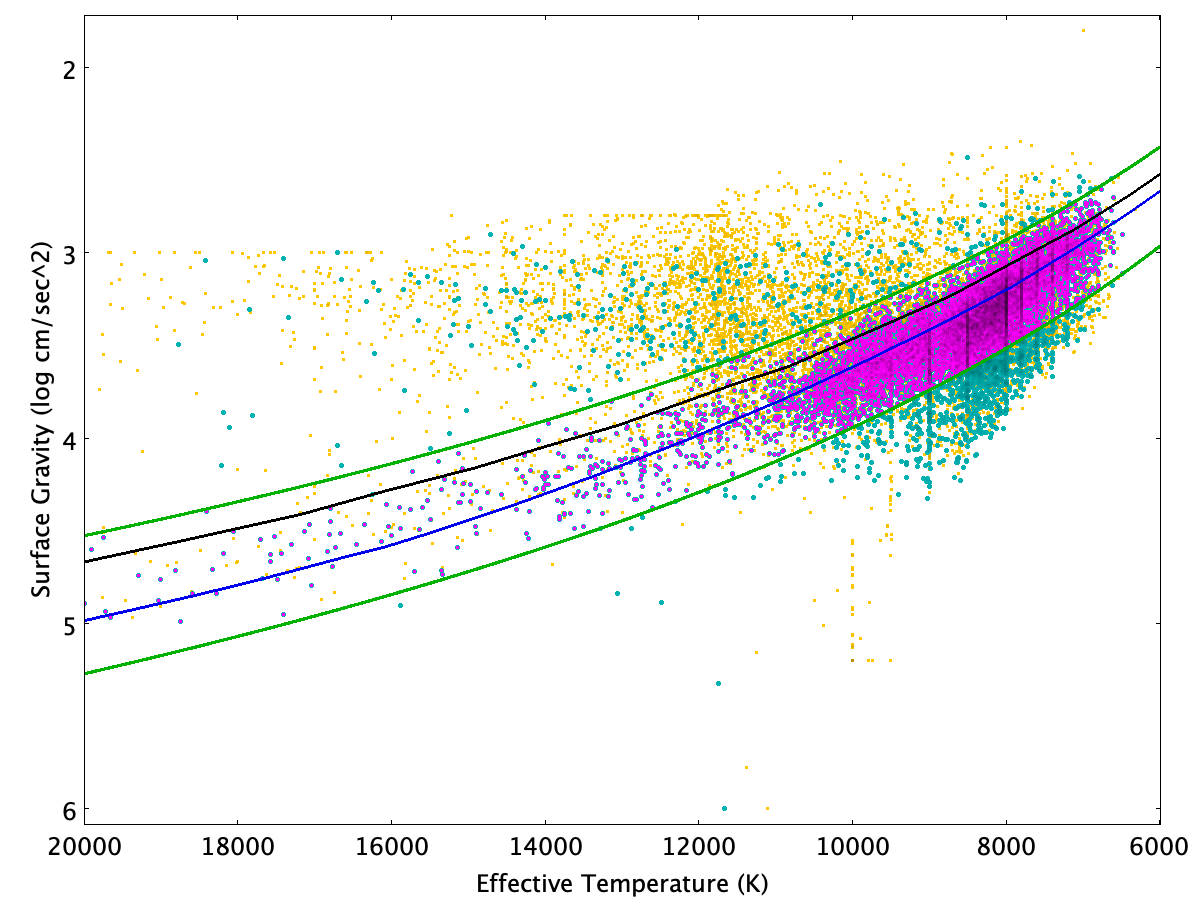}
  \includegraphics[width=\hsize]{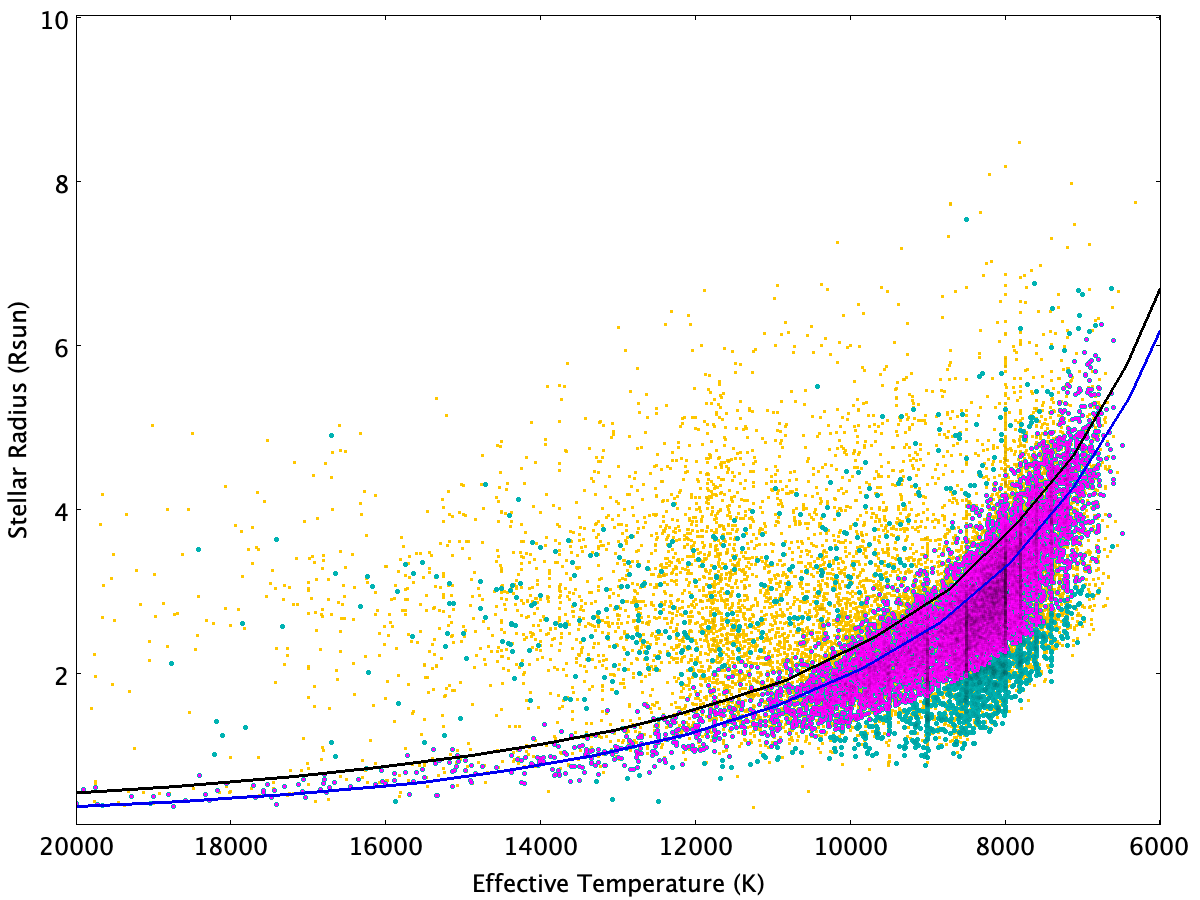}
  \includegraphics[width=\hsize]{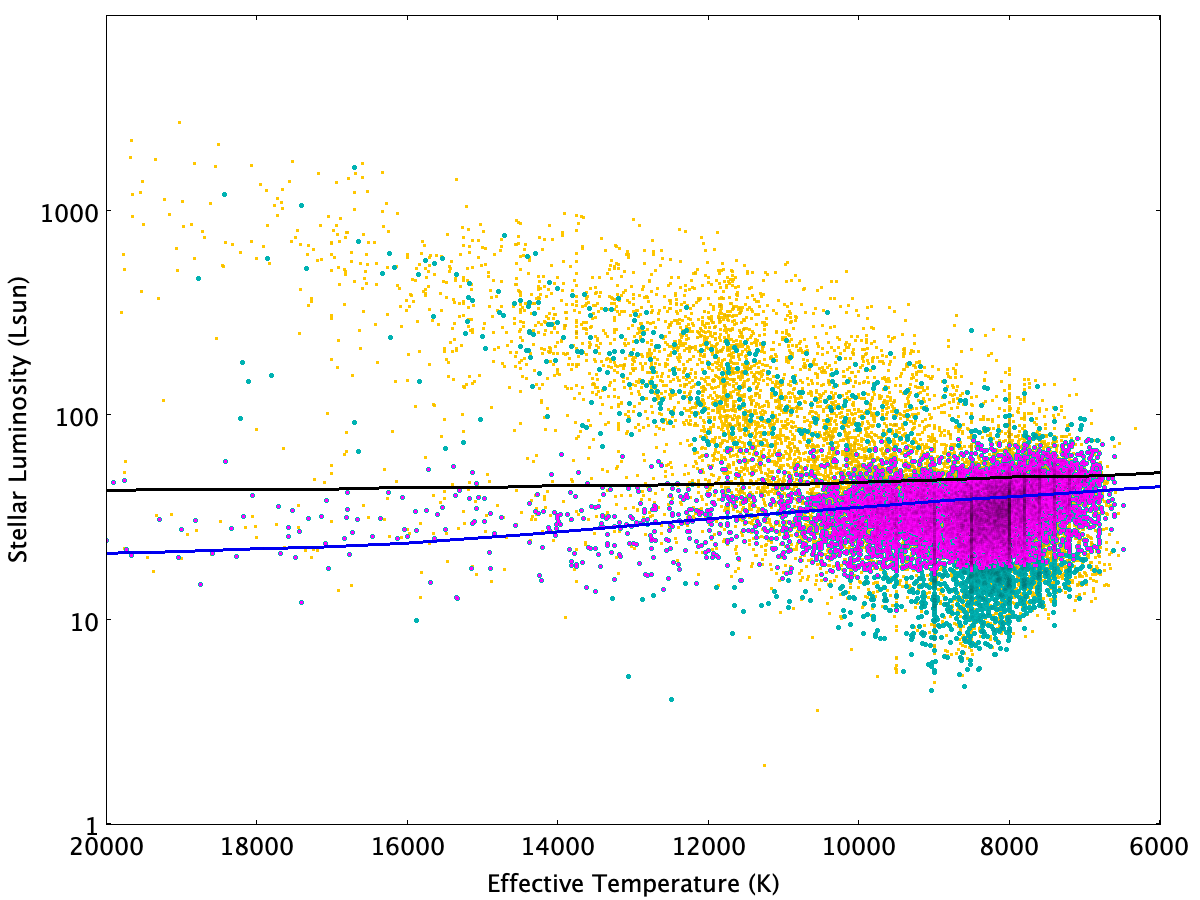}
 \caption{\textit{Upper panel}: $T_\mathrm{eff}$ versus $\log g$ as calculated from the 22,335 BHB candidate objects where an SED calculation was made (yellow dots), 10,604 SEDs where the full frequency range is available (cyan dots) of which 9,172 (magenta dots) lie within the horizontal branch region as marked by the green lines defined by Eq.\ \ref{eq:tahb_cutoff} and \ref{eq:zahb_cutoff}. The BaSTI ZAHB (blue line) and TAHB (black line) are also shown.  \textit{Middle panel}: $T_\mathrm{eff}$ versus $R$. \textit{Lower Panel}: $T_\mathrm{eff}$ versus $L$ for the same subsets.}
  \label{galex_dorman_hb}
  \end{figure}

When plotting the $T_\mathrm{eff}$ against the $\log g$ (see Figure~\ref{galex_dorman_hb}) we found that 9,172 of the 10,604 objects with full photometric range data plotted within the horizontal branch region, defined by the cut-off lines
\begin{equation}
\label{eq:tahb_cutoff}
    \mathrm{cut_{lower}} = 4.4157  \log_{10}(T_\mathrm{eff}) - 13.720,
\end{equation}
\vspace{-20pt}
\begin{equation}
\label{eq:zahb_cutoff}
    \mathrm{cut_{upper}} = 4.0150  \log_{10}(T_\mathrm{eff}) - 12.742. 
\end{equation}
These lines are based on the ZAHB and TAHB isochrones from BaSTI models given in \citet{pietrinferni04}, but somewhat widened to include the coherent cloud of BHBs. 
This gave an estimated contamination level of 14\%. We confirmed these selection criteria by plotting the same BHB candidates in $T_\mathrm{eff}$ versus $R$ space, as well as the $T_\mathrm{eff}$ versus $L$ space where they also occupy a coherent region around the ZAHB and TAHB isochrones (see Figure~\ref{galex_dorman_hb}).

\subsection{Contamination estimation from spectral analysis}

Spectra were acquired for BHB candidates with a {\em Gaia} DR3 apparent $G$ magnitude of brighter than 11.0\,mag at the Ond\v{r}ejov Observatory in the Czech Republic over a 18 month period. The spectra were acquired using the Perek 2m telescope through an Echelle spectrograph with a resolving power of $R$ = 51,600 around H$\alpha$. 
The resulting spectra were reduced using the IRAF\footnote{IRAF is distributed by the National Optical Astronomy Observatories, which are operated by the Association of Universities for Research in Astronomy, Inc., under cooperative agreement with the National Science Foundation.} \citep{tody86,tody93} package and a dedicated semi-automatic pipeline \cite{cabezas23} which incorporates a full range of standard procedures for echelle spectra reduction, including bias correction, flat-fielding, wavelength calibration, heliocentric calibration, and continuum normalization.

\begin{figure*}[t]
    \centering
    
    \begin{subfigure}[t]{0.49\textwidth}
        \centering
        \includegraphics[width=1\textwidth]{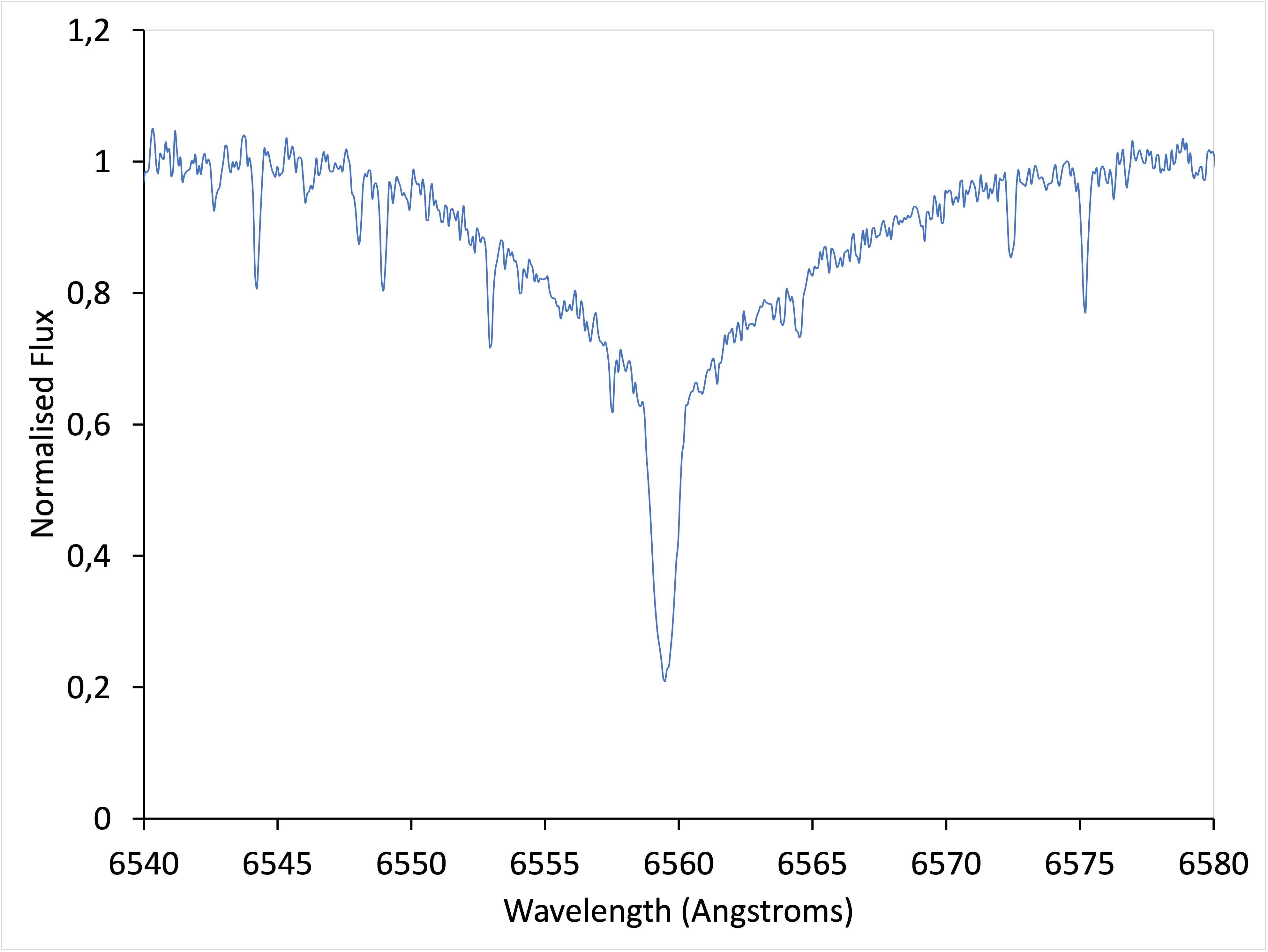}
        \caption{TYC 1738-745-1 (H$\alpha$)}
        \label{TYC_1738_745_1_Halpha}
    \end{subfigure}
    \hfill
    \begin{subfigure}[t]{0.49\textwidth}
        \centering
        \includegraphics[width=1\textwidth]{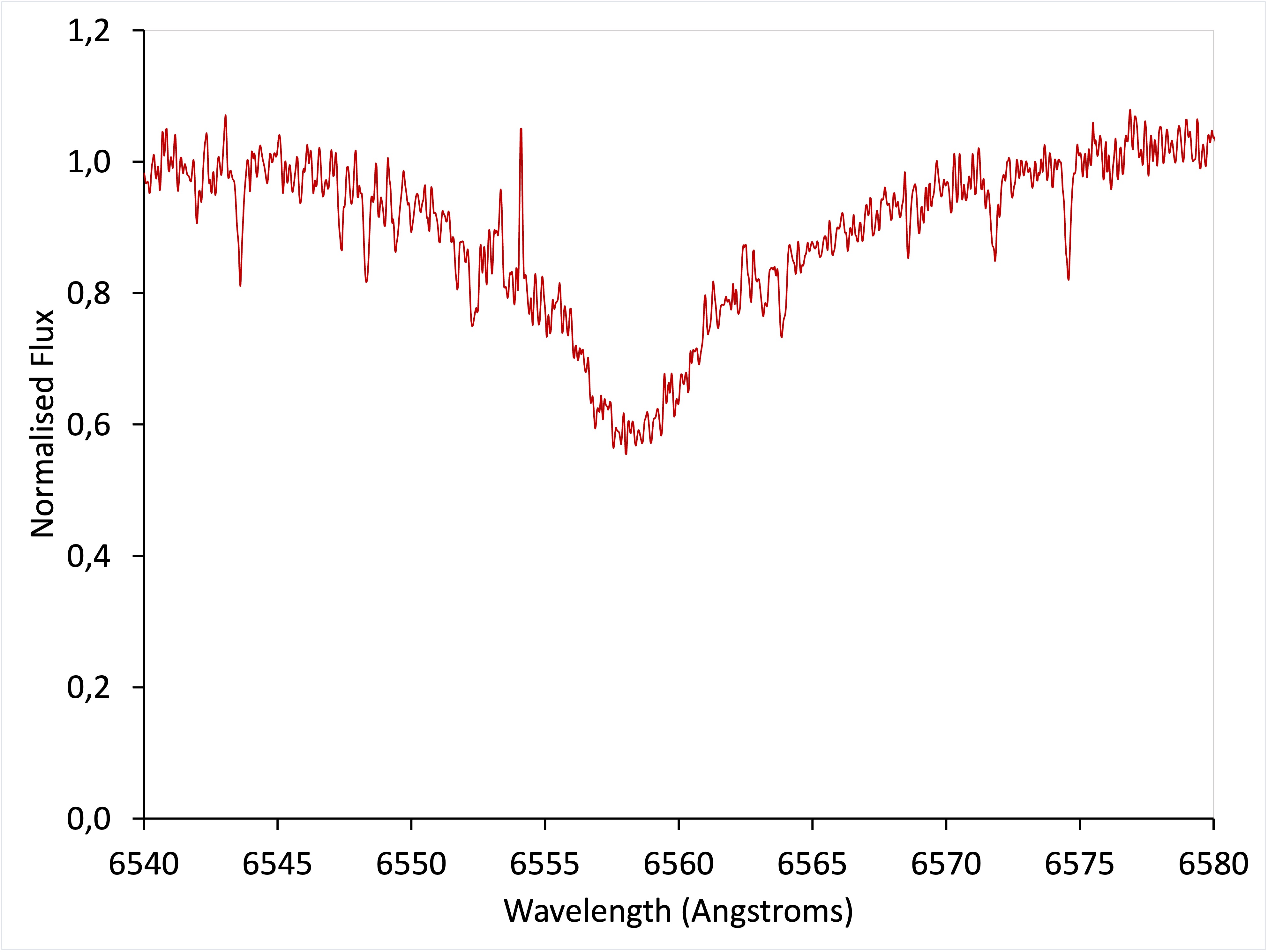}
        \caption{TYC 3655-1179-1 (H$\alpha$)}
        \label{TYC_3655_1179_1_Halpha}
    \end{subfigure}

    \begin{subfigure}[t]{0.49\textwidth}
        \centering
        \includegraphics[width=1\textwidth]{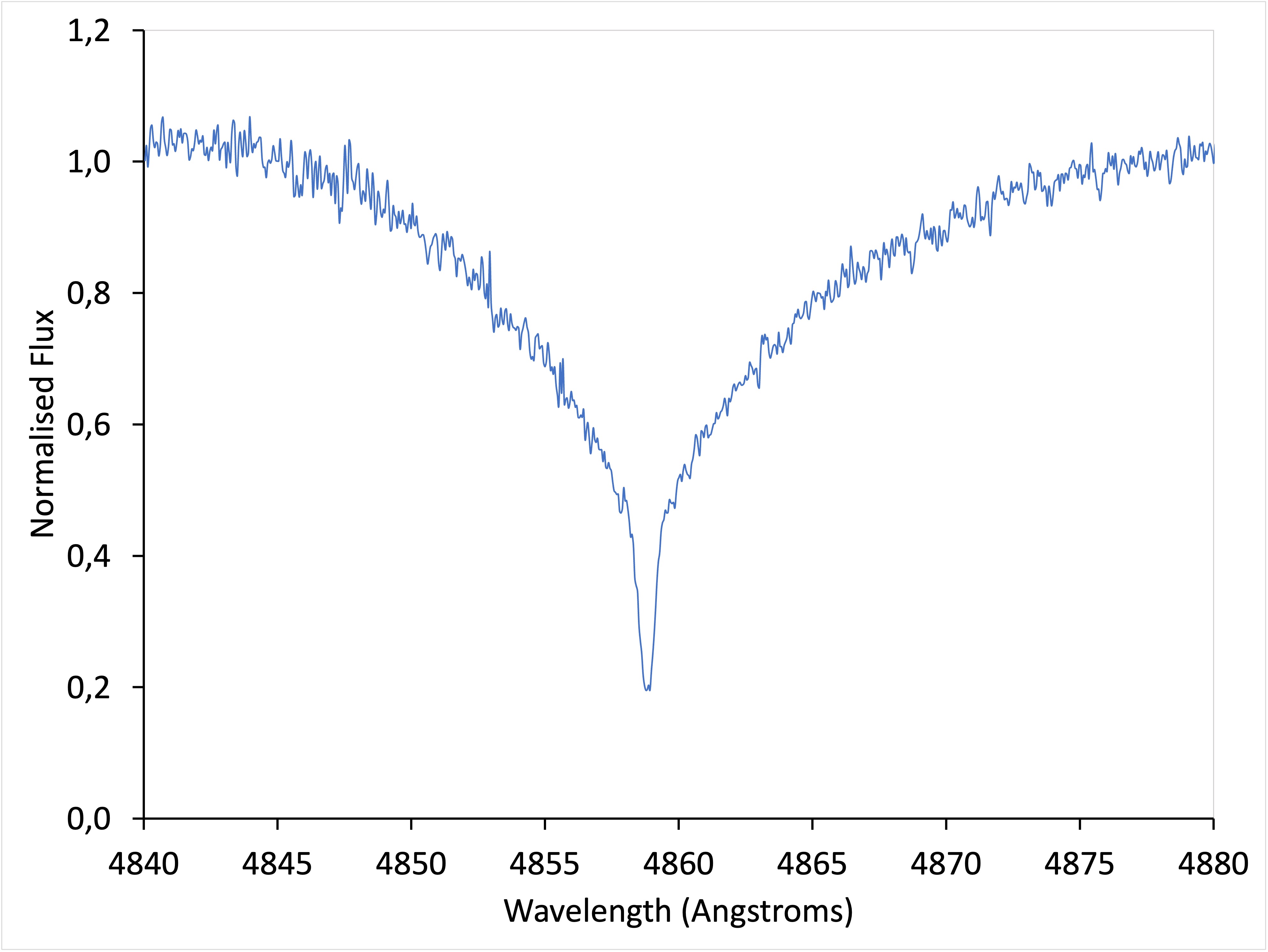}
        \caption{TYC 1738-745-1 (H$\beta$)}
        \label{TYC_1738_745_1_Hbeta}
    \end{subfigure}
    \hfill
    \begin{subfigure}[t]{0.49\textwidth}
        \centering
        \includegraphics[width=1\textwidth]{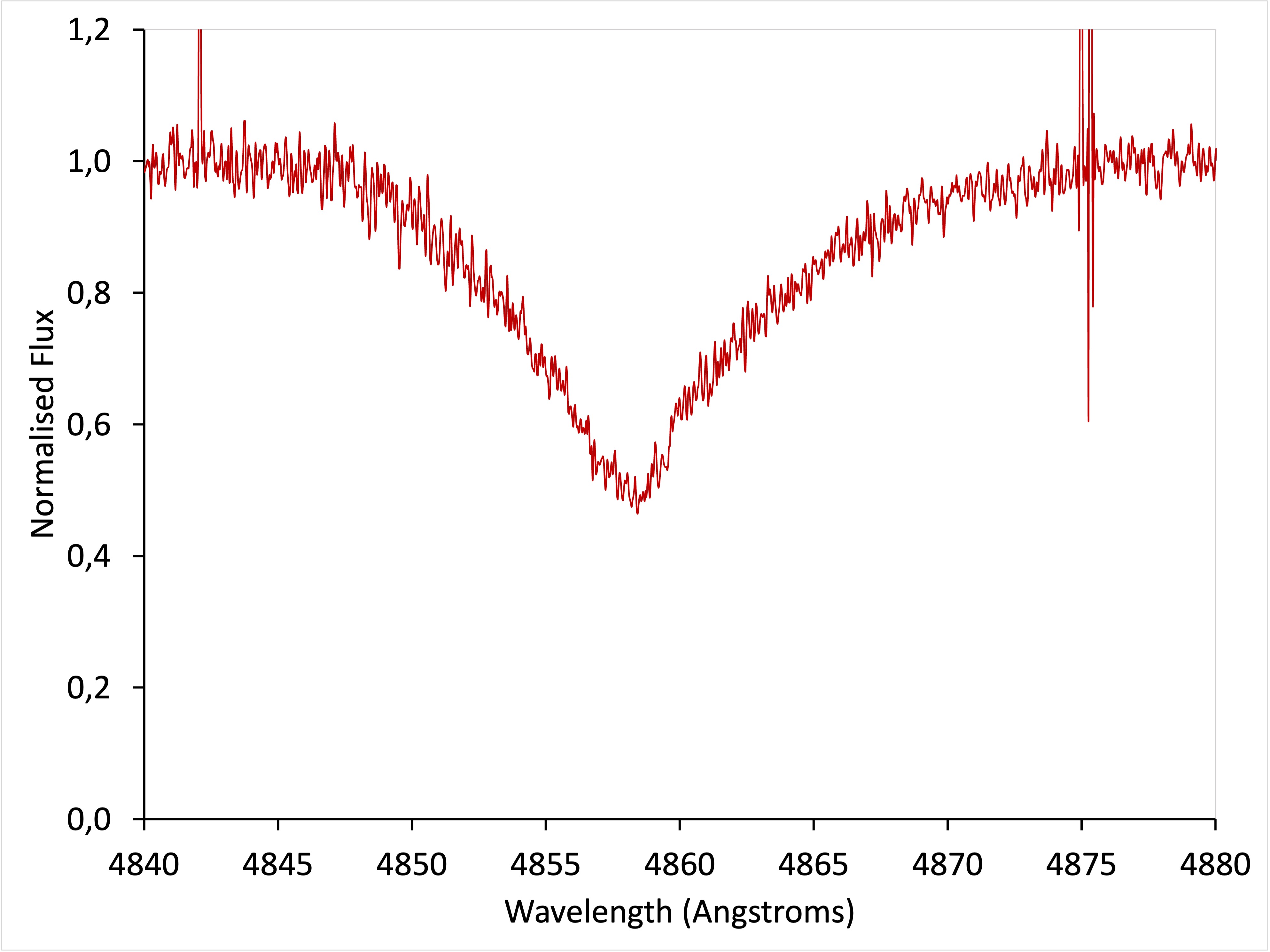}
        \caption{TYC 3655-1179-1 (H$\beta$)}
        \label{TYC_3655_1179_1_Hbeta}
    \end{subfigure}
     \caption{Reduced spectra for a BHB star TYC 1738-745-1 (left panels) and a high proper motion main-sequence star TYC 3655-1179-1 (right panels) for H$\alpha$ (upper panels) and H$\beta$ (lower panels).}
  \label{spectra}
\end{figure*}

The criteria used to differentiate between spectra from BHB stars and from main-sequence stars were the width and the depth of the Balmer series hydrogen spectral lines in a similar fashion to \citet{xue08}. Examples of BHB and main-sequence spectra are shown in Figure~\ref{spectra}. Over an 18 month period we acquired spectra for 69 BHB candidates of which 6 were found to be main-sequence stars giving a $9\pm 3$\,\%  contamination level, similar to that found with the SED analyses (Section \ref{sect:cont_sed}) albeit with a far smaller sample.

\section{Sky Coverage, magnitude, and distance}
\label{sect:sky}

\begin{figure}
  \centering
  \includegraphics[width=\hsize]{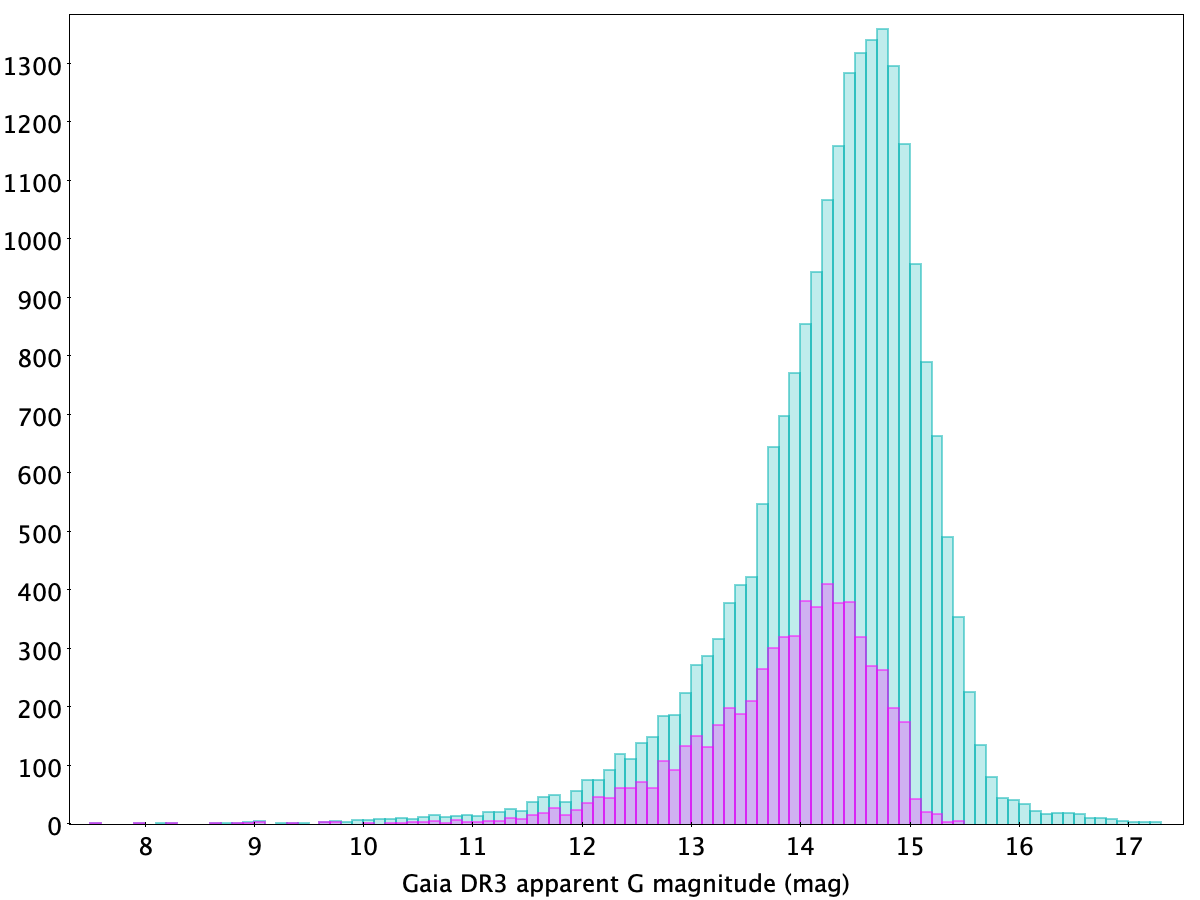}
 \caption{Comparison of the \citet{culpan21} {\em Gaia} DR3 apparent magnitude distribution of the BHB candidate objects from the SED based catalogue (magenta) and the  {\em Gaia} DR3 based catalogue (cyan). The peak values of these 2 histograms are found at similar values indicating that both catalogues are investigating similar volumes. Finding additional BHB candidates in the same volume indicates increased completeness of the new catalogue.}
  \label{app_mag_histogram}
  \end{figure}

The extended catalogue of BHB candidate stars presented here covers a similar apparent magnitude range and hence a similar distance range as the \citet{culpan21} predecessor but contains 32\% more BHB candidates. Thus, the level of completeness of the catalogue has been improved (see Figure~\ref{app_mag_histogram}).

  \begin{figure}
  \centering
  \includegraphics[width=\hsize]{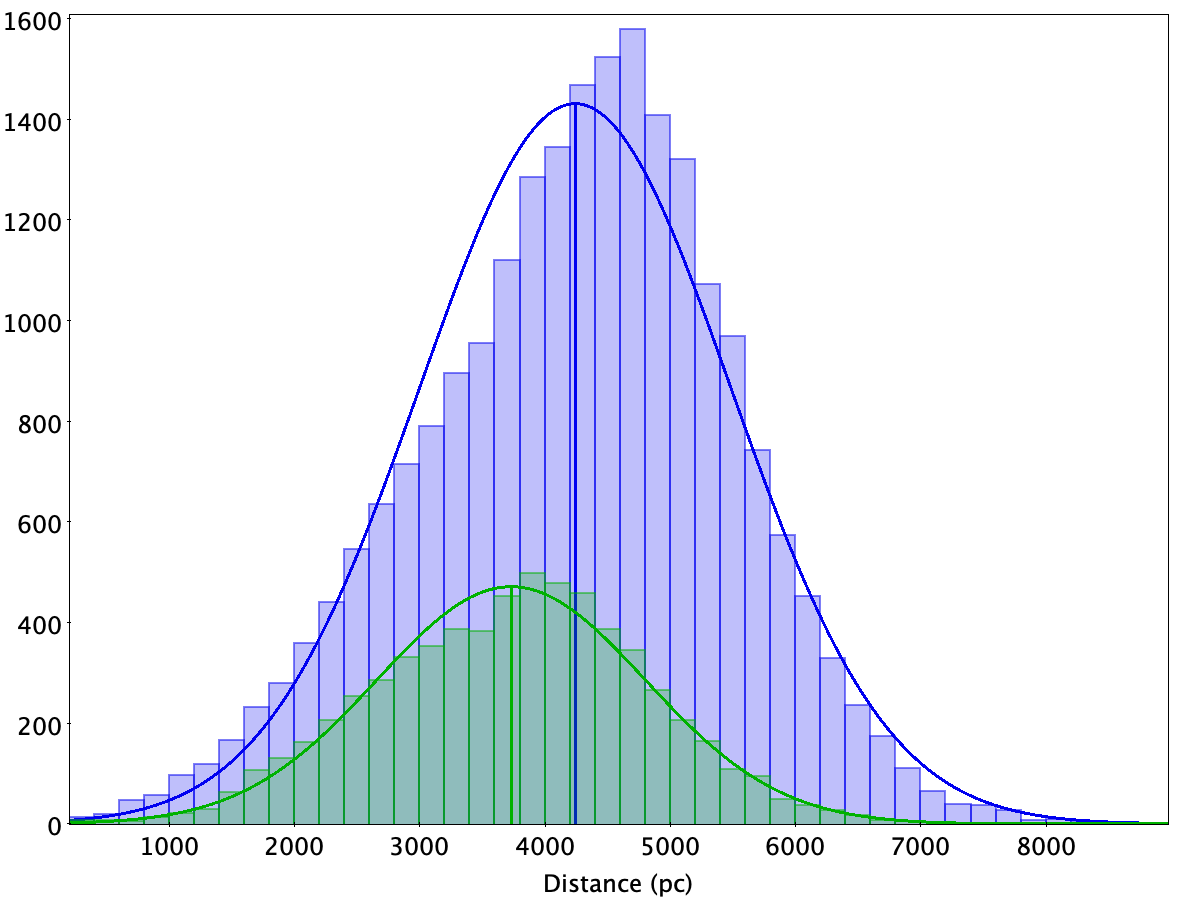}
 \caption{Histogram of the geometric distance \citep{bailer21} to BHB candidates from the Gaia DR3 based catalogue (blue) and the SED based catalogue of BHB candidates with stellar parameters (green).}
  \label{bhb_distance}
  \end{figure}

The catalogue of BHB candidates with stellar parameters calculated from SEDs using full photometry range data contains 9,172 stars with a sky coverage as shown in Figure~\ref{all_sky}, a distance range of 235\,pc to 7,435\,pc as shown in Figure~\ref{bhb_distance} calculated from the zero-point corrected parallax \citep{bailer21} and an apparent magnitude range of 7.6 mag to 15.5 mag as shown in Figure~\ref{app_mag_histogram}.

\section{Summary and Conclusions}
\label{sect:summary}

The {\em Gaia} DR3 catalogue of BHB star candidates is an update to the parallax selected catalogue presented in \citet{culpan21}. This latest version of the catalogue has improved coverage at the hotter end of the BHB at $T_\mathrm{eff}$ greater than 10,500\,K. The data quality of {\em Gaia} DR3 permits less restrictive criteria being applied in regions of high apparent stellar density allowing more BHB candidates to be found. The Final Parallax Selection in the \citet{culpan21} catalogue had 16,794 candidate objects. The extended catalogue presented here (see Figure~\ref{all_sky} and Appendix A) contains 22,335 BHB candidates. This represents an increase of 32\%. 

We have used photometric data from 66 large scale surveys to generate synthetic SEDs and associated stellar parameters for a large number of BHB candidates where no spectra are available. The use of such a rich dataset has allowed reliable stellar parameters to be computed. The modelled SEDs have then been used to refine the {\em Gaia} DR3 CMD selection criteria and to demonstrate contamination levels of less than 12\% in the catalogue.

\begin{figure}
  \centering
  \includegraphics[width=\hsize]{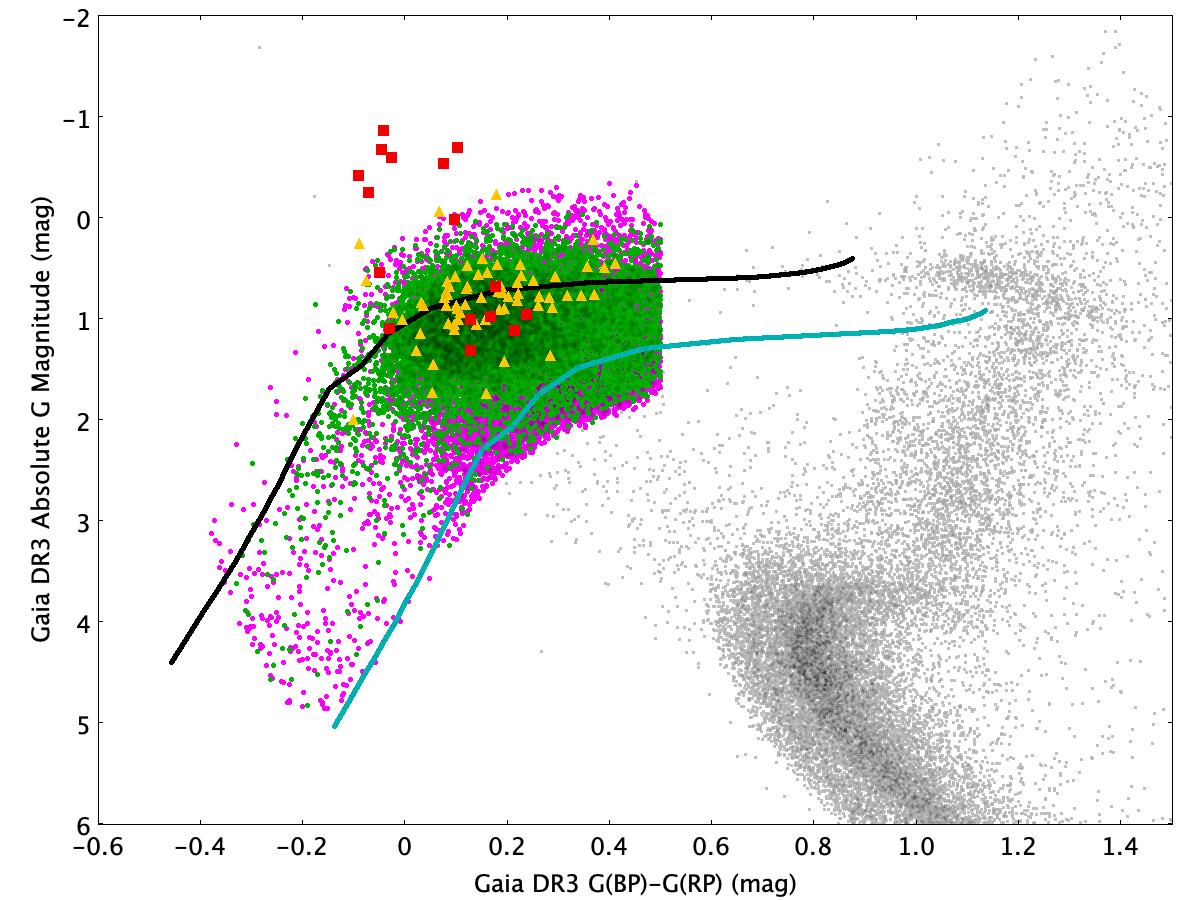}
 \caption{The SED based catalogue of Galactic Halo BHB candidate stars. {\em Gaia} DR3 CMD showing objects with a tangential velocity greater than 145 km\,s$^{-1}$ (grey dots). The {\em Gaia} DR3 based catalogue of BHB stars (magenta dots), the SED confirmed BHB stars (green dots), spectrally confirmed BHB stars (yellow  triangles), spectrally confirmed main-sequence stars (red squares). Zero mag extinction (black line) and 0.2 mag extinction (cyan line) theoretical tracks are superimposed.}
  \label{results}
  \end{figure}

This contamination level has also been observationally confirmed using spectra acquired in the Ond\v{r}ejov Observatory in the Czech Republic (see Figure~\ref{results}).

We have demonstrated that the use of theoretically interpolated SEDs is a powerful tool in constructing a richer, more complete and less contaminated catalogue than was done in the past.

This increase of 32\% in the number of BHB candidate stars compared to \citet{culpan21} but within the same volume and apparent magnitude range, represents another step in generating a more complete full-sky catalogue of BHB stars in the inner Galactic Halo.

We then generated the SEDs as described above and used the results to extinction correct the {\em Gaia} DR3 $(G_{BP} - G_{RP})$ and $G_\mathrm{abs}$ values. This was then used to define the lower absolute magnitude limit of the {\em Gaia} DR3 CMD BHB candidate selection region.

A further iteration was made using these newly defined {\em Gaia} DR3 CMD selection criteria and 22,335 BHB candidates (see Figure~\ref{results}) were found. Of these candidates 10,222 (46\%) had photometric data available that supported the generation of stellar- and atmospheric parameters from SEDs. The stellar- and atmospheric parameters generated indicate that 9,172 (41\%) of these objects lie in the parameter space expected for BHBs.

\begin{acknowledgements}

We thank Andreas Irrgang (University of Erlangen-N\"uenberg) for amending the ATLAS12 code and developing the SED fitting tool.  

I.P. acknowledges funding by the UK’s Science and Technology Facilities Council (STFC), grant ST/T000406/1, and from a Warwick Astrophysics prize post-doctoral fellowship made possible thanks to a generous philanthropic donation. 

B.K, and M.C. acknowledge the support from the Grant Agency of the Czech Republic (GA\v CR 22-34467S) and RVO:67985815 and the Stellar Department of the Astronomical Institute of the Czech Academy of Science.

This research made use of TOPCAT, an interactive graphical viewer and editor for tabular data Taylor (\cite{taylor05}). This research made use of the SIMBAD database, operated at CDS, Strasbourg, France; the VizieR catalogue access tool, CDS, Strasbourg, France. Some of the data presented in this paper were obtained from the Mikulski Archive for Space Telescopes (MAST). STScI is operated by the Association of Universities for Research in Astronomy, Inc., under NASA contract NAS5-26555. Support for MAST for non-HST data is provided by the NASA Office of Space Science via grant NNX13AC07G and by other grants and contracts. This research has made use of the services of the ESO Science Archive Facility.

This work has made use of data from the European Space Agency (ESA) mission {\it Gaia} (https://www.cosmos.esa.int/gaia), processed by the {\it Gaia} Data Processing and Analysis Consortium (DPAC, https://www.cosmos.esa.int/web/gaia/dpac/consortium). Funding for the DPAC has been provided by national institutions, in particular the institutions participating in the {\it Gaia} Multilateral Agreement.

This work has made use of BaSTI web tools.

\end{acknowledgements}

\clearpage
\begin{onecolumn}
\begin{appendix}

\section{Gaia DR3 catalogues of BHB candidate stars}

\begin{longtable}{llll}
\caption{\label{table:A1} Catalogue columns. }\\
\hline\hline
\noalign{\smallskip}
Column & Format & Description & Unit \\
\noalign{\smallskip}
\hline
\noalign{\smallskip}
source\_id & I19 & Gaia EDR3 source identifier & - \\
ra & F10.6 & Gaia EDR3 Right ascension & deg \\
dec & F10.6 & Gaia EDR3 Declination & deg \\
l & F10.6 & Galactic longitude & deg \\
b & F10.6 & Galactic latitude & deg \\
parallax & F8.4 & Gaia parallax $\bar{\omega}$ & mas \\
parallax\_error & F8.4 & Error on parallax & mas \\
abs\_g\_mag & F8.4 & Absolute magnitude in G-band & mag \\
abs\_g\_mag\_zpc & F8.4 & Absolute magnitude in G-band from zpc parallax & mag \\
phot\_g\_mean\_mag & F6.3 & Gaia apparent G magnitude & mag \\
bp\_rp & F6.3 & Gaia $G_{BP}$ - $G_{RP}$ magnitude & mag \\
phot\_bp\_rp\_excess\_factor\_corrected & F6.3 & Corrected Gaia $G_{BP}$ - $G_{RP}$ excess factor & - \\
phot\_bp\_rp\_excess\_factor\_corrected\_error & F6.3 & Error on corrected Gaia $G_{BP}$ - $G_{RP}$ excess factor & - \\
pmra & F7.3 & Gaia proper motion $\mu_{\rm \alpha}\cos{\rm \delta}$ & ${\rm mas\,yr^{-1}}$ \\
pmra\_error & F7.3 & Error on pmra & ${\rm mas\,yr^{-1}}$ \\
pmdec & F7.3 & Gaia proper motion $\mu_{\rm \delta}$ & ${\rm mas\,yr^{-1}}$ \\
pmdec\_error & F7.3 & Error on pmdec & ${\rm mas\,yr^{-1}}$ \\
pm & F7.3 & Gaia proper motion $\mu$ & ${\rm mas\,yr^{-1}}$ \\
pm\_error & F7.3 & Error on proper motion & ${\rm mas\,yr^{-1}}$ \\
vt & F7.3 & Tangential velocity & $(km.s^{-1})$ \\
vt\_error & F7.3 & Error on tangential velocity & $(km.s^{-1})$ \\
excess\_flux\_error & F7.3 & Excess flux error & - \\
zeropoint & F7.3 & parallax zero point & mas  \\
parallax\_zpc & F8.4 & zero point corrected parallax & mas \\
within\_5\_arcsec & F8.4 & Gaia objects within 5 arcsec of BHB candidate & - \\
5\_arcsec\_flux\ & F8.4 & Total G flux from within 5 arcsec radius of BHB candidate & counts \\
candidate\_flux\_fraction & F8.4 & Fraction of 5 arcsec flux that comes from BHB candidate & - \\
SED\_flag & Boolean & Flag for SED calculation & - \\
SED\_logtheta & f8.4 & Angular diameter $log(\Theta)$ & - \\
SED\_logtheta\_minus\_error & f8.4 & Angular diameter minus error $log(\Theta)_{min}$ & - \\
SED\_logtheta\_plus\_error & f8.4 & Angular diameter plus error $log(\Theta)_{max}$ & - \\
SED\_e\_44m55 & f8.4 & Reddening & mag \\
SED\_e\_44m55\_minus\_error & f8.4 & Reddening minus error & mag \\
SED\_e\_44m55\_plus\_error & f8.4 & Reddening plus error & mag \\
SED\_teff & f8.4 & Effective temperature $T_{eff}$ & K \\
SED\_teff\_minus\_error & f8.4 & Effective temperature minus error $(T_{eff})_{min}$ & K \\
SED\_teff\_plus\_error & f8.4 & Effective temperature plus error $(T_{eff})_{max}$ & K \\
SED\_logg & f8.4 & Surface gravity $log(g)$ & $log(cm/sec^2)$ \\
SED\_logg\_minus\_error & f8.4 & Surface gravity minus error $log(g)_{min}$ & $log(cm/sec^2)$ \\
SED\_logg\_plus\_error & f8.4 & Surface gravity plus error $log(g)_{max}$ & $log(cm/sec^2)$ \\
SED\_z & f8.4 & Metallicity & - \\
SED\_r & f8.4 & Median stellar radius from SED calculation & $R_\odot$ \\
SED\_r\_minus\_error & f8.4 & Stellar radius minus error from SED calculation & $R_\odot$ \\
SED\_r\_plus\_error & f8.4 & Stellar radius plus error from SED calculation & $R_\odot$ \\
SED\_m & f8.4 & Median stellar mass from SED calculation & $M_\odot$ \\
SED\_m\_minus\_error & f8.4 & Stellar mass minus error from SED calculation & $M_\odot$ \\
SED\_m\_plus\_error & f8.4 & Stellar mass plus error from SED calculation & $M_\odot$ \\
SED\_l & f8.4 & Median stellar luminosity from SED calculation & $L_\odot$ \\
SED\_l\_minus\_error & f8.4 & Stellar luminosity minus error from SED calculation & $L_\odot$ \\
SED\_l\_plus\_error & f8.4 & Stellar luminosity plus error from SED calculation & $L_\odot$ \\

\noalign{\smallskip}
\hline\hline
\end{longtable}
\begin{landscape}
\section{Photometric surveys used for synthetic SED generation}
\label{appendix:photometry}

\begin{longtable}{lrl}
\caption{\label{table:B1} 66 Surveys used in SED calculations. }\\
\hline\hline
\noalign{\smallskip}
Survey & Data count & Citation \\
\noalign{\smallskip}
\hline
\noalign{\smallskip}
The DENIS Database & 42502 & \citet{denis05} \\
Dark Energy Camera Plane Survey (DECaPS) DR1 & 8762 & \citet{schlafly18} \\
DECam Local Volume Exploration Survey (DELVE) DR2 & 29080 & \citet{drlica22} \\
Dark Energy Survey (DES) DR2 & 6201 & \citet{abbott21} \\
Far Ultraviolet Spectroscopic Explorer (FUSE) & 39 & \citet{dixon09} \\
Gaia Data Release 3 BP/RP low-resolution spectral data & 329732 & \citet{deangeli22} \\
Hubble Source Catalog (V1 and V2) & 16 & \citet{whitmore16} \\
Tycho-2 Catalogue & 2509 & \citet{hog00} \\
Hipparcos Catalogue & 38 & \citet{vanleeuwen07} \\
Yale/San Juan Southern Proper Motion (SPM) Catalog 4 & 35970 & \citet{girard11} \\
Gaia EDR3 & 92827 & \citet{riello21} \\
Homogeneous Means in the UBV System & 104 & \citet{mermilliod06} \\
The Geneva Photometry Catalogue & 47 & \citet{rufener99} \\
uvby-beta Catalogue & 233 & \citet{hauck98} \\
Stellar Photometry in Johnson's 11-color system & 1 & \citet{ducati02} \\
2 Micron All-Sky Survey - Catalog of Point Sources & 90454 & \citet{skrutskie06} \\
Beijing-Arizona-Taiwan-Connecticut (BATC) Large Field Multi-Color Sky Survey & 186 & \citet{xu05} \\
IRSF Magellanic Clouds Point Source Catalog & 135 & \citet{kato08} \\
Galactic Legacy Infrared Mid-Plane Survey Extraordinaire (GLIMPSE) & 102 & \citet{spitzer09}\\
Spitzer Survey of the Large Magellanic Cloud & 250 & \citet{meixner06} \\
UKIDSS-DR6 Galactic Plane Survey & 282 & \citet{ukidss12} \\
UKIDSS-DR9 LAS, GCS and DXS Surveys & 6421 & \citet{lawrence13} \\
AllWISE Data Release & 58617 & \citet{cutri13} \\
Revised catalog of GALEX UV sources & 29991 & \citet{bianchi17} \\
AAVSO Photometric All Sky Survey (APASS) DR9 & 114422 & \citet{henden15} \\
Swift/UVOT Serendipitous Source Catalog & 799 & \citet{yershov14} \\
KiDS-ESO-DR3 multi-band source catalog & 194 & \citet{dejong15} \\
Pan-STARRS release 1 (PS1) Survey & 70115 & \citet{chambers16} \\
VLT Survey Telescope ATLAS & 7680 & \citet{shanks15} \\
XMM-OM Serendipitous Source Survey Catalogue & 608 & \citet{page12} \\
The Dark Energy Survey (DES): Data Release 1 & 5387 & \citet{abbott18} \\
The band-merged unWISE Catalog & 60584 & \citet{schlafly19} \\
CatWISE2020 catalog & 59057 & \citet{marocco21} \\
Catalogue of stellar UV fluxes & 12 & \citet{thompson78} \\
UBVRIJKLMNH Photoelectric Catalogue & 11 & \citet{morel78} \\
Spectroscopically Identified Hot Subdwarf Stars & 23 & \citet{kilkenny88} \\
UBV(RI)cHalpha photometry in omega Cen & 588 & \citet{bellini09} \\
South Galactic cap MCT blue objects & 1 & \citet{lamontagne00} \\
Magellanic Clouds Photometric Survey: The LMC & 182 & \citet{zaritsky04} \\
UBVRI Standard Stars & 1 & \citet{landolt07} \\
Omega Centauri Spitzer photometry & 491 & \citet{boyer08} \\
Extended Kepler-INT Survey & 1522 & \citet{greiss12} \\
JK photometry of 12 galactic globular clusters & 6 & \citet{cohen15} \\
OGLE LMC BVI photometry & 12 & \citet{udalski00} \\
UBVI photometry in NGC 6752 & 9 & \citet{kravtsov16} \\
UBV photometry of metal-weak candidates & 15 & \citet{norris99} \\
The ZTF catalog of periodic variable stars & 152 & \citet{chen20} \\
Spitzer Kepler Survey (SpiKeS) catalog & 200 & \citet{werner21} \\
Edinburgh-Cape Blue Object Survey. Zone 1 & 30 & \citet{kilkenny97} \\
Edinburgh-Cape Blue Object Survey. III. & 42 & \citet{odonoghue13} \\
Edinburgh-Cape Blue Object survey. IV & 28 & \citet{kilkenny15} \\
Edinburgh-Cape Blue Object survey. V & 15 & \citet{kilkenny16} \\
UBVRI photometry in 48 globular clusters & 1917 & \citet{stetson19} \\
Javalambre Photometric Local Universe Survey Data Release DR3 & 16533 & \citet{vonmarttens24} \\
Pan-STARRS Data Release 2 & 67667 & \citet{flewelling18} \\
Survey of the Magellanic Stellar History (SMASH) DR2 & 2182 & \citet{nidever17} \\
Southern Photometric Local Universe (S-PLUS) Survey DR3 & 12934 & \citet{mendesdeoliveira19} \\
SkyMapper Southern Survey: DR2 & 124543 & \citet{onken19} \\
SDSS Photometric Catalogue, Release 12 & 30601 & \citet{alam15} \\
IGAPS. merged IPHAS and UVEX of northern Galactic plane & 1562 & \citet{monguio20} \\
Visible and Infrared Survey Telescope for Astronomy DR6 & 41589 & \citet{mcmahon21} \\
Final Merged Log of International Ultraviolet Explorer Observations & 1636 & \citet{nasa85} \\
VISTA Deep Extragalactic Observations (VIDEO) Survey DR5 & 33 & \citet{jarvis13} \\
VISTA Kilo-degree Infrared Galaxy Public Survey (VIKING) DR4 & 1384 & \citet{edge13} \\
VISTA Magellanic Survey (VMC) catalog DR4 & 98 & \citet{cioni11} \\
VISTA Variable in the Via Lactea Survey (VVV) DR4 & 1539 & \citet{minniti23} \\

\noalign{\smallskip}
\hline\
\end{longtable}
\begin{tiny}
Acknowledgements: This project used public archival data from the Dark Energy Survey (DES). Funding for the DES Projects has been provided by the U.S. Department of Energy, the U.S. National Science Foundation, the Ministry of Science and Education of Spain, the Science and Technology FacilitiesCouncil of the United Kingdom, the Higher Education Funding Council for England, the National Center for Supercomputing Applications at the University of Illinois at Urbana-Champaign, the Kavli Institute of Cosmological Physics at the University of Chicago, the Center for Cosmology and Astro-Particle Physics at the Ohio State University, the Mitchell Institute for Fundamental Physics and Astronomy at Texas A\&M University, Financiadora de Estudos e Projetos, Funda{\c c}{\~a}o Carlos Chagas Filho de Amparo {\`a} Pesquisa do Estado do Rio de Janeiro, Conselho Nacional de Desenvolvimento Cient{\'i}fico e Tecnol{\'o}gico and the Minist{\'e}rio da Ci{\^e}ncia, Tecnologia e Inova{\c c}{\~a}o, the Deutsche Forschungsgemeinschaft, and the Collaborating Institutions in the Dark Energy Survey.
The Collaborating Institutions are Argonne National Laboratory, the University of California at Santa Cruz, the University of Cambridge, Centro de Investigaciones Energ{\'e}ticas, Medioambientales y Tecnol{\'o}gicas-Madrid, the University of Chicago, University College London, the DES-Brazil Consortium, the University of Edinburgh, the Eidgen{\"o}ssische Technische Hochschule (ETH) Z{\"u}rich,  Fermi National Accelerator Laboratory, the University of Illinois at Urbana-Champaign, the Institut de Ci{\`e}ncies de l'Espai (IEEC/CSIC), the Institut de F{\'i}sica d'Altes Energies, Lawrence Berkeley National Laboratory, the Ludwig-Maximilians Universit{\"a}t M{\"u}nchen and the associated Excellence Cluster Universe, the University of Michigan, the National Optical Astronomy Observatory, the University of Nottingham, The Ohio State University, the OzDES Membership Consortium, the University of Pennsylvania, the University of Portsmouth, SLAC National Accelerator Laboratory, Stanford University, the University of Sussex, and Texas A\&M University.
Based in part on observations at Cerro Tololo Inter-American Observatory, National Optical Astronomy Observatory, which is operated by the Association of Universities for Research in Astronomy (AURA) under a cooperative agreement with the National Science Foundation.

This work presents results from the European Space Agency (ESA) space mission Gaia. Gaia data are being processed by the Gaia Data Processing and Analysis Consortium (DPAC). Funding for the DPAC is provided by national institutions, in particular the institutions participating in the Gaia MultiLateral Agreement (MLA).

Based on observations made with the NASA/ESA Hubble Space Telescope, and obtained from the Hubble Legacy Archive, which is a collaboration between the Space Telescope Science Institute (STScI/NASA), the Space Telescope European Coordinating Facility (ST-ECF/ESAC/ESA) and the Canadian Astronomy Data Centre (CADC/NRC/CSA). 

The UKIDSS project is defined in Lawrence et al (2007). UKIDSS uses the UKIRT Wide Field Camera (WFCAM; Casali et al, 2007). The photometric system is described in Hewett et al (2006), and the calibration is described in Hodgkin et al. (2009). The pipeline processing and science archive are described in Irwin et al (2009, in prep) and Hambly et al (2008).

This publication makes use of data products from the Wide-field Infrared Survey Explorer, which is a joint project of the University of California, Los Angeles, and the Jet Propulsion Laboratory/California Institute of Technology, funded by the National Aeronautics and Space Administration.

Based on data products from observations made with ESO Telescopes at the La Silla Paranal Observatory under programme IDs 177.A-3016, 177.A-3017 and 177.A-3018, and on data products produced by Target/OmegaCEN, INAF-OACN, INAF-OAPD and the KiDS production team, on behalf of the KiDS consortium. OmegaCEN and the KiDS production team acknowledge support by NOVA and NWO-M grants. Members of INAF-OAPD and INAF-OACN also acknowledge the support from the Department of Physics \& Astronomy of the University of Padova, and of the Department of Physics of Univ. Federico II (Naples).

Based on observations obtained as part of the VISTA Hemisphere Survey, ESO Progam, 179.A-2010.

This publication has made use of data from the VIKING survey from VISTA at the ESO Paranal Observatory, programme ID 179.A-2004. Data processing has been contributed by the VISTA Data Flow System at CASU, Cambridge and WFAU, Edinburgh.

\end{tiny}
\end{landscape}

\section{ADQL query for BHB candidates}

The following query can be used in the https://gea.esac.esa.int/archive/ website to get a file containing 30,228 BHB candidates from the {\em Gaia} DR3 data set:

\begin{longtable}{l}
\noalign{\smallskip}
$\verb!SELECT *!$ \\
$\verb!FROM gaiadr3.gaia_source!$ \\
$\verb!WHERE parallax_over_error >= 5!$ \\
$\verb!AND parallax > 0!$ \\
$\verb!AND (4.74 / parallax * pm) >= 145!$ \\
$\verb!AND phot_g_mean_mag + 5 + 5*log10(parallax/1000) < 138.07*power(bp_rp,6) - 153.85*power(bp_rp,5)!$  \\
$\verb!  - 40.727*power(bp_rp,4) + 73.368*power(bp_rp,3) - 7.4054*power(bp_rp,2) - 9.5575*bp_rp + 3.8459!$ \\
$\verb!AND phot_g_mean_mag + 5 + 5*log10(parallax/1000) > -3.2382*power(bp_rp,3) + 7.1259*power(bp_rp,2)!$ \\
$\verb!  - 3.583*bp_rp - 0.2!$ \\
$\verb!AND bp_rp < 0.5 AND bp_rp > -0.4!$ \\
\end{longtable}

\end{appendix}

\end{onecolumn}

\end{document}